\documentclass[11pt,a4paper]{article}
\usepackage{jcappub}

\usepackage{epsfig,psfrag}

\begin{document}
 
\title{Thermal decoupling of WIMPs from first principles}

\author[a]{Torsten Bringmann}
\emailAdd{torsten.bringmann@fys.uio.no}
\affiliation[a]{Department of Physics, University of Oslo, Box 1048, NO-0316 Oslo, Norway}

\author[b]{Stefan Hofmann}
\emailAdd{stefan.hofmann@physik.lmu.de}
\affiliation[b]{Arnold Sommerfeld Center for Theoretical Physics, Ludwig-Maximilians-Universit\"at, Theresienstra\ss e 37, 80333 Munich, Germany}

\date{Apr 9, 2016}

\abstract{
 Weakly interacting massive particles (WIMPs) are arguably the most natural DM candidates from a particle physics point of view.
 After their number density has frozen out in the early universe, determining their relic density today, WIMPs are still kept very close to thermal equilibrium by scattering events with standard model particles. The complete decoupling from the thermal bath happens as late as around $\sim1-10\,$MeV, thereby setting an important cosmological scale that can directly be translated into a small-scale cutoff of the spectrum of matter density fluctuations. We present here a full analytic treatment of the kinetic decoupling process from first principles. This allows an exact determination of the decoupling scale, for an arbitrary WIMP candidate and any scattering amplitude. As an application, we consider the situation of the lightest supersymmetric particle as well as the lightest Kaluza-Klein particle that arises in theories with universal extra dimensions; furthermore, we show that our formalism can also easily be applied to, e.g., the evolution of the non-relativistic electrons into the recombination regime. Finally, we comment on the impacts for the smallest gravitationally bound structures and the prospects for the indirect detection of dark matter.
}
\maketitle

\newcommand{\be}{\begin{equation}}
\newcommand{\ee}{\end{equation}}
\newcommand{\bea}{\begin{eqnarray}}
\newcommand{\eea}{\end{eqnarray}}
\newcommand{\kref}[1]{(\ref{#1})}
\newcommand{\B}{{B^{(1)}}}

\section{Introduction}
Based on the simplest inflationary scenario, a flat cosmological model with dark energy and cold dark matter as dominant contingents 
in the cosmic energy budget fits the spectrum of observed temperature anisotropies imprinted in the cosmic microwave background  radiation (CMB) \cite{wmap}.
Assuming that the primordial fluctuations are adiabatic with a power law spectrum, the temperature anisotropy data {\sl alone} 
indicates a significant amount of non-brayonic cold dark matter (CDM);
analysis of the anisotropies finds that the relative matter density $\omega_{\rm m} \equiv \Omega_{\rm m} h^2 = 0.127^{+0.007}_{-0.013}$
is roughly six times larger than the relative baryon density.
This is consistent with a diversity of other cosmological observations, e.g. the analysis of galaxy clustering as observed by SDSS \cite{Eisenstein:2005su}
and 2dFGRS \cite{Cole:2005sx}, improvements in the lensing data \cite{Hoekstra:2002xs}, as well as improvements
in small-scale CMB measurements,e.g. \cite{Leitch:2004gd}.

Arguably the best studied class of CDM particle candidates are weakly interacting massive particles, or WIMPs (for reviews on particle DM, see \cite{DMreviews}). Equipped with masses and typical interaction strengths set by the electroweak scale, they naturally occur 
in many extensions to the standard model of particle physics (SM), e.g. in its supersymmetric versions or in models with space-like extra dimensions.
Thermally produced in the early universe, and provided their stability, WIMPs automatically acquire a relic abundance that is of the right order of magnitude to account for the observed matter density. 
Another well-studied CDM particle candidate is the axion. The essential
difference between WIMPs and axions is that the former have been kept in local thermal equilibrium prior to some epoch in the cosmological evolution while the latter have not. This is because by the time  
the axion potential developed a minimum at around the QCD phase transition, and the axions thus became massive, any interactions were already heavily suppressed by the Peccei-Quinn scale. 
In this article, we will focus solely on WIMPs and often use the abbreviation CDM as a synonym for WIMPs.

It has long been a widely held belief that the process of structure formation is completely independent of the CDM particle nature; CDM density perturbations, in particular, should
only be effected by their gravitational interaction and bear no visible imprint of particle interactions.
In this case, however, cosmological perturbation theory 
predicts a monotonously growing support for structure formation
on ever smaller scales. More precisely, the spectrum of matter density perturbations grows logarithmically with the comoving wavenumber that characterizes 
the perturbation mode \cite{Green:2003un,Green:2005fa}. This behaviour has recently been confirmed yet another time, down to the resolution of a numerical simulation with unprecedented accuracy \cite{Diemand:2006ik}. As a consequence, a regularization scheme would be required to render linear perturbation theory well-defined,
e.g.~a cut-off in the wavenumber that corresponds to some resolution limit. From cosmological perturbation theory we know that  gravitational interactions alone
cannot give rise to a fundamental cut-off scale, so it has to be introduced by hand. On the other hand, it is rather problematic to impose such an ad hoc small-scale cut-off since, in hierarchical structure formation, small scale perturbations enter the nonlinear regime first and structures on larger scales form only later via mergers and accretion. 

In Ref.~\cite{Hofmann:2001bi} it was shown that CDM interactions with the radiation component give rise to a natural cut-off that can be calculated from first principles,
i.e. from the interaction theory underlying a specific dark matter candidate. The authors of \cite{Berezinsky:2003vn} confirmed the results of 
\cite{Hofmann:2001bi}. 
The (physical) cut-off is set by viscous processes around the kinetic decoupling of CDM from radiation. 
Close to kinetic decoupling, CDM density perturbations receive in-medium modifications that give rise to a dispersion relation
with a non-vanishing imaginary contribution and thus to absorption.
Subsequently, free streaming of CDM suppresses further the power of structure formation on small scales.
In that way, collisional and collisionless damping processes give rise to a fundamental cut-off in the spectrum of linear 
CDM perturbations \cite{Green:2003un,Green:2005fa}
and thus to a characteristic feature that clearly depends on the CDM microphysics. 
It was later shown numerically \cite{Loeb:2005pm} that acoustic oscillations have to be taken into account at kinetic decoupling,
however, the order of magnitude of the cut-off scale is correctly given in \cite{Green:2003un,Green:2005fa}.
Most recently, the author of \cite{Bertschinger:2006nq} succeeded in including acoustic oscillations before and during
kinetic decoupling, the viscous coupling during kinetic decoupling and free streaming in a consistent kinetic treatment.
Close to this physical cut-off scale, the power spectrum develops a maximum. Hence, the cut-off scale characterizes 
density perturbations that typically enter the non-linear regime first and therefore give rise to the first purely gravitationally
bound structures.

Provided that the present day dark matter density distribution retains traces of these typical first halos, the physical cut-off scale
may thus have important observational consequences.
While it is certainly true that the present CDM density profile is independent of the particle nature of CDM on super-kpc scales, this is highly
questionable on sub-pc scales. Let us therefore stress that WIMP detection experiments actually probe the dark matter density profile on sub-mpc scales:  
for a given WIMP candidate, the prospects for a direct detection (via elastic scattering processes with the detector nuclei) 
depend on the effective reaction volume which is given by the detector size and the path the detector travels
during data taking.  Indirect detection experiments, on the other hand, aim at measuring flux densities from dark matter annihilation products, which depend quadratically on the dark matter density contrast in the observed region.
Substructures thus enhance the fluxes and therefore the prospects for indirect detection, probing
the dark matter density profile on sub-galactic scales.

It is really the temperature at kinetic decoupling that provides the interface, at least in principal, between the interaction theory underlying a specific
CDM candidate and CDM substructures that are probed by direct and indirect dark matter searches. 
In this article we present
a complete and analytic description of kinetic decoupling of WIMPs from the local heat bath, without assuming a 
specific WIMP candidate.  In Section \ref{sec_dec} we introduce important time scales that characterize the non-equilibrium state
of WIMPs, discuss on a very qualitative level chemical and kinetic decoupling of WIMPs, before we derive a precise description of
the kinetic decoupling process. We introduce a generalized WIMP temperature (which characterizes the WIMP distribution
even in the kinetic stage and agrees with the heat bath temperature before kinetic decoupling) and derive a master equation for the WIMP temperature. 
Section \ref{sec_ex} is devoted to various examples: as a first and known example we quantify the deviation of the electron temperature from the
photon temperature during the recombination regime. Much more relevant for our purposes is the next example, the kinetic decoupling of 
bino-like CDM from SM fermions. As a last example we consider Kaluza-Klein dark matter.
Section \ref{sec_clumps} provides a summary of the processes that constitute the small-scale cut-off in the linear matter power spectrum; here, we also shortly review what is known about the total mass fraction in and survival probability of the smallest gravitionally bound structures that correspond to this scale.
In section \ref{sec_inddet} we discuss prospects for indirect dark matter detection experiments in light of small-scale structure formation. 
Section \ref{sec_conc}, finally, concludes. All technical details and derivations are presented in the three appendices: In \ref{app_kd}, we solve the Boltzmann equation for the collision integral, as determined in \ref{app_collint}, that describes the scattering processes of relevance to our discussion; in \ref{app_LKPscatter}, we present the LKP scattering amplitude and use this opportunity to demonstrate the applicability of our formalism even in the case of $s$-channel resonances of the scattering amplitude.

\section{Thermal relics and their decoupling from the heat bath}
\label{sec_dec}
In this section, we first introduce various time scales that allow to characterize different stages in the evolution of WIMPs. 
While a thorough understanding of the various characteristic time scales is not required
for the later computation of the kinetic decoupling temperature, 
the presentation mainly aims at clarifying the importance of the relaxation time for WIMPs close to local thermal equilibrium.
 We then discuss chemical 
and kinetic decoupling of WIMPs from the local heat bath
on a very qualitative level, before proceeding to a precise description of kinetic decoupling -- which is the main concern of this article.

\subsection{Characteristic time scales}

The most detailed statistical description of WIMPs is given by the $N$-particle phase space density, with $N$
denoting their number in a comoving Volume $V$. Around kinetic decoupling,
the interaction radius $r_{\rm int}$ of WIMPs (defined as the maximal distance between a WIMP and SM -- or, rather, heat bath -- particles
over which interactions can proceed) is much smaller than the mean distance  between the particles, i.e.
$N r_{\rm int}^{\; 3}/V \ll 1$. This simple inequality allows the introduction of different time scales that are important
for the classical kinetic description of dark matter. The first time scale is called {\sl collision time} and defined by 
$\tau_{\rm c} \equiv r_{\rm int} /\langle v\rangle$, with $\langle v\rangle$ denoting the mean dark matter velocity. 
A second time scale is provided by the time between two subsequent collisions,
$\tau_{\rm f}\equiv l_{\rm f}/\langle v\rangle$, where $l_{\rm f}$ is the mean free path of dark matter particles. The third
and for our considerations most important time scale is the {\sl relaxation time} $\tau_{\rm r}$. 
It marks the time during which local thermal equilibrium is established in a comoving volume that
contains many dark matter particles, but that is still small compared to $V$. The last time scale $\tau_{\rm eq}$
is given by the time during which global equilibrium in $V$ is established. 

From elementary kinetic theory it follows that the mean free path can be approximated by $l_{\rm f}/V^{1/3} \sim (1/N) (V^{1/3}/r_{\rm int})^2$, 
hence $\tau_{\rm c} \ll \tau_{\rm f}$. Since CDM particles with weak interactions
need many collisions in order to maintain local thermal equilibrium, we also have $\tau_{\rm f} \ll \tau_{\rm r}$. Finally, one evidently has $\tau_{\rm r} \ll \tau_{\rm eq}$.
 To summarize, we get the following chain of inequalities for the characteristic time scales introduced above:
\be
\tau_{\rm c} \ll \tau_{\rm f} \ll \tau_{\rm r} \ll  \tau_{\rm eq}\,.
\ee

For a description of processes on time scales $\delta t \ll \tau_{\rm c}$, the full $N$-particle phase space density is required,
and a reduced description in terms of some $\tilde{N}<N$-particle phase space distribution is impossible.
A reduced kinetic description is, however, available for processes with $\tau_{\rm c} \ll \delta t \ll \tau_{\rm r}$. The WIMPs
can then adequately be described by a one-particle distribution that determines the probability distribution for the 
locations and the momenta of the CDM particles. 
For processes on time scales $\tau_{\rm r} \ll \delta t \ll \tau_{\rm eq}$, the system is in local thermal equilibrium and
further simplifications apply. During this stage, the WIMPs are fully 
described by the local number, momentum and kinetic energy density, which are given as the zeroth, first and second
moments, reypectively, of the one-particle distribution with respect to the WIMP momentum. This is the underlying reason that allows to describe such a system
as a perfect or almost perfect fluid. 
 
 \subsection{Scales of decoupling}
 
Initially, the WIMP system is (almost) in local chemical and thermal equilibrium, with chemical equilibrium
provided by the detailed balance between WIMP annihilation and production processes, and
local thermal equilibrium maintained by elastic scattering processes with SM particles in the heat bath. 
Typically, chemical and thermal equilibrium are protected by the aforementioned processes for temperatures
above the WIMP particle mass, $T\gtrsim M_\chi$. 

As the Universe expands, the temperature decreases and the WIMP momenta become redshifted. At $T\sim M_\chi$
detailed balance is destroyed: the lighter species, dominating the energy budget, 
can no longer produce WIMPs, while the reverse channel is still open. As a consequence, 
between the end of detailed balance at $T\sim M_\chi$ and chemical decoupling at $T=T_{\rm cd}$, the total number
of WIMP particles decreases. Chemical decoupling happens roughly when the 
WIMP annihilation rate drops below the Hubble rate $H$, and, as a consequence,
annihilation processes cease to proceed and the relic dark matter abundance is fixed.
This is expected to happen since the relevant
target density in the annihilation processes is provided by the WIMP number density, which is Boltzmann
suppressed for temperatures $T< M_\chi$. 
For instance, in the case of WIMPs the chemical decoupling scale is typically given by $T_{\rm cd}\sim  M_\chi/25$. 
Hence, the total number of WIMPs is conserved for $T\gtrsim M_\chi$ -- dynamically protected
by detailed balance, decreases during $M_\chi/25\lesssim T\lesssim M_\chi$, and again protected for $T\lesssim M_\chi/25$ -- basically
by the volume expansion.\footnote{Strictly speaking, the relic abundance is protected until density perturbations
generate density contrasts that again allow for WIMP annihilations.}
Note that the dependence of the chemical decoupling scale $T_{\rm cd}$ on the WIMP annihilation rate (and thus
on the underlying interaction theory) can be
completely eliminated using the value for the relative WIMP density as inferred e.g. from the temperature anisotropy data.

After chemical decoupling elastic scattering processes still keep WIMPs in local thermal equilibrium. It is very important
to realize that even though dark matter is now a chemically distinct particle species, it is still a constituent of the local heat bath. 
The relevant target density for elastic scattering processes is provided by the number density of relativistic SM
particles and thus only decreasing as $T^3$. Eventually, also the elastic scattering rate $\Gamma_{\rm el}$ 
drops below the Hubble rate
and elastic scattering processes cease. However, even before last scattering, local thermal equilibrium can not be maintained.
The relaxation processes within the WIMP system are weakened by the Hubble expansion. 
Consequently, the relevant time scales are the relaxation time $\tau_{\rm r}$ (intrinsic) and the Hubble time $H^{-1}$ (extrinsic).

Before we will present a way to analytically determine the decoupling scale in the next Section,
let us now roughly estimate the relaxation time for WIMPs as follows (see also \cite{Hofmann:2001bi}): our system consists of a weakly interacting
particle species with mass $M_\chi$ and a local heat bath characterized by a temperature $T\ll M_\chi$, so $M_\chi$ is the
dominant energy scale. These conditions lead to a small relative momentum transfer during 
 single elastic scattering process, $\delta p/p \sim (T/ M_\chi)^{1/2} \ll 1$. 
We assume that $\delta p/p$ is constant during the relaxation process. 
After $N_{\rm coll}$ collisions, the WIMP momentum will thus have changed by $\delta p/p (N_{\rm coll})^{1/2}$. 
Ideally, we want to know, after how many scattering processes 
local thermal equilibrium has been established again. For a rough estimate, without solving the
appropriate kinetic theory, we have to be more modest and estimate instead the number
of collisions required to change the WIMP momentum significantly. Clearly, the answer
is $N_{\rm coll} \sim M_\chi/T\gg 1$. 
The relaxation time can be estimated as
$\tau_{\rm r}\sim N_{\rm coll}/\Gamma_{\rm el}\sim (m/T) 1/\sigma_{\rm scale}$,
where we assume that the scale for the total elastic scattering process is set by
$\sigma_{\rm scale} \equiv (G_{\rm F} m_{\rm W}^{\; 2}) M_\chi^{\; 2}/m_{\rm Z}^{\; 4}$, with $G_{\rm F}$ denoting Fermi's constant, $m_{\rm W}$ and $m_{\rm Z}$ are the masses of the
charged and neutral weak gauge bosons, respectively.
Kinetic decoupling occurs approximately at the temperature $T_{\rm kd}$, for which $\tau_{\rm r} (T_{\rm kd}) = H^{-1}(T_{\rm kd})$. 
In the toy model under
considerations, we thus expect
$T_{\rm kd} \sim  M_\chi^{\; 1/4} m_{\rm Z}/M_{\rm Pl}^{\; 1/4} (G_{\rm F} m_{\rm W}^{\; 2})^{1/2}\sim {\cal O}(1-10)$ MeV.


\subsection{A precise calculation of the kinetic decoupling process}

Around kinetic decoupling WIMPs pass from the hydrodynamical to the kinetic stage. The description 
of WIMPs changes accordingly: while the hydrodynamical stage describes the evolution of fluid variables that are
given as expectation values with respect to the one-particle phase space distribution, the kinetic
stage is concerned with the evolution of the one-particle phase space distribution $f({\bf p})$ itself. 

A more precise determination of the decoupling temperature requires the introduction of a temperature parameter 
$T_{\chi}$ for dark matter, which we choose to define as
\begin{equation}
\label{def}
T_\chi
\equiv
\frac{2}{3}
\left\langle\frac{{\bf p}^2}{2M_\chi}\right\rangle
\; , \hspace{0.5cm} {\rm with} \; 
\left\langle{\cal O}({\bf p})\right\rangle
\equiv 
\frac{1}{n_\chi} \int \frac{{\rm d}^3p}{(2\pi)^3}\; {\cal O}({\bf p}) f({\bf p})
\; .
\end{equation}
Here, $n_\chi$ denotes the WIMP number density, ${\cal O}$ is a microscopic observable
and $f$ is the WIMP phase-space distribution.  The parameter $T_\chi$ that we introduced above agrees with the heat bath temperature for $T>T_{\rm kd}$ 
and can be viewed as a generalized WIMP temperature for $T<T_{\rm kd}$, characterizing the non-equilibrium  
state of WIMPs.

The dynamics of the one-particle phase space distribution, governed by the Boltzmann equation, is nontrivial during the 
kinetic decoupling process (kinetic stage).
In the case of dark matter weakly coupled to radiation the main complication arises from the
coupling of slow processes (non-relativistic propagation of dark matter) and fast processes (relativistic propagation of radiation) 
in the collision integral, which is enforced via the explicit energy momentum
conservation constraining the phase space available for elastic scattering processes.

Around kinetic decoupling, the WIMP mass $M_\chi$ is by far the largest energy scale involved. In particular, the kinetic
WIMP energy, as well as the average momentum
transfer during the elastic scattering processes, is tiny compared to $M_\chi$. This allows to factorize slow and fast processes in the collision integral by utilizing 
the Born-Oppenheimer approximation and to expand the collision integral in 
${\bf p}^2/M_\chi^{\; 2}$.
Instead of solving the Boltzmann equation for $f$,
we can then solve its second moment and derive the following process equation for $T_\chi(T)$ in Appendix A:
\begin{equation}
\label{master}
\frac{{\rm d}T_\chi}{{\rm d}T} - \left[ 2 + a \left(\frac{T}{M_\chi}\right)^{n+2}\right]\frac{T_\chi}{T}
=
- a \left(\frac{T}{M_\chi}\right)^{n+2}
\, .
\end{equation}  
Here, $n$ is defined by the low-energy behaviour of the scattering amplitude at zero momentum transfer, $\left|\mathcal{M}\right|^2_{t=0}\propto (\omega/M_\chi)^n$, where $\omega$ is the energy of the SM scattering partner; $a$
is a numerical constant that is defined in (A.23).  
The uniquely defined solution to Eq.~(\ref{master}) with the correct asymptotic behavior, $T_\chi\stackrel{T\rightarrow\infty}{\rightarrow}T$, is given by:
\be
  \label{TCDMmain}
    T_\chi=T\left\{1-\frac{z^{1/(n+2)}}{n+2} \exp[z]\, \Gamma\left[-(n+2)^{-1},z\right]\right\}_{z=\frac{a}{n+2}\left(\frac{T}{M_\chi}\right)^{n+2}} \, .
\ee
In the limit $T\rightarrow 0$, we have $T_\chi\sim T^2/M_\chi$. Since, as we shall see, kinetic decoupling occurs on a rather short time scale, a suitable choice for a \emph{definition of the kinetic decoupling temperature} is obtained by matching the two above-mentioned asymptotic solutions to Eq.~(\ref{master}), resulting in: 
\be
  \label{tdecmain}
  \frac{T_\mathrm{kd}}{M_\chi}\equiv \left(\left(\frac{a}{n+2}\right)^{1/(n+2)}\Gamma\left[\frac{n+1}{n+2}\right]\right)^{-1}\,.
\ee

For a more detailed discussion of the process equation (\ref{master}), we refer to \ref{app_kd} and the following Section, where we 
consider some examples of particular interest and determine the exact values of the decoupling temperature in these cases. However, from \kref{adef}, let us already now remark that we generically expect it to be of the order of
\be
\label{Tkd_OM}
\frac{T_\mathrm{kd}}{M_\chi}\sim\left(\alpha^{-2}M_\chi/M_\mathrm{Pl}\right)^{1/(n+2)}\,,
\ee
where $\alpha$ is the effective coupling constant that appears in the scattering amplitude. For example, we have $\alpha=\alpha_\mathrm{em}$, $n=0$ in the case of recombination (Thompson scattering) and $\alpha=\alpha_Y$, $n=2$ for the thermal decoupling of Binos or LKP photons.  From Eq.~\kref{Tkd_OM} we can see that, indeed, the WIMP microphysics enters in a crucial way in the determination of the kinetic decoupling scale and thus -- see Section \ref{sec_clumps} -- in the numerical value for the cutoff in the spectrum of CDM density perturbations.

\section{Examples}
\label{sec_ex}

In this section, we apply the formalism developed in Appendix A and B, as reviewed in the proceeding Section, for illustrative purposes first to the decoupling of electrons from radiation during
the recombination era. In a second step, we turn to something more related to the main focus of this article and discuss the kinetic decoupling of typical WIMP DM candidates, i.e.~bino-like and LKP dark matter, in particular.

\subsection{Recombination}

After the (chemical) decoupling of electrons and positrons at a temperature of around $T\sim1$~MeV, and the subsequent annihilation of $e^+e^-$ pairs, the photons were still interacting with protons and the remaining electrons, mainly through Compton scattering with the latter. Such a plasma of interacting relativistic and non-relativistic particles cannot, strictly speaking, be kept in thermal equilibrium in an expanding universe. On the other hand, CMB observations show a black-body spectrum to an accuracy of about $10^{-4}$ \cite{Fixsen:1996nj}. As a first application of our formalism, we will now quantify the expected deviation from a thermal distribution at the time when the universe became transparent  and the CMB photons that we observe today were released. For a similar treatment of the transition into the recombination regime, see, e.g. \cite{Bernstein}.

At zero momentum transfer and to lowest order in the photon energy, the amplitude for Thompson scattering is given by
\be
  \left|\mathcal{M}\right|^2_{t=0}=64\pi^2\alpha^2_\mathrm{em}\,.
\ee
From our general solution \kref{TCDMmain}, the temperature of the non-relativistic electrons therefore evolves as 
\bea
  \label{Te}
  T_{e}&=&T\left\{1-\frac{z^{1/2}}{2} \exp[z]\, \Gamma\left[-1/2,z\right]\right\}_{z=\frac{a}{2}\left(\frac{T}{m_e}\right)^{2}}\\
       &=&T\sqrt{\pi z}\exp[z]\left\{1-\mathrm{Erf}\left[\sqrt{z}\right]\right\}_{z=\frac{a}{2}\left(\frac{T}{m_e}\right)^{2}}\,,
\eea
where
\be
   a \stackrel{(\ref{adef})}{=} \sqrt{\frac{10}{(2\pi)^9\cdot 3.36}}\,128\pi^2\alpha^2_\mathrm{em}N_3^- \frac{M_\mathrm{Pl}}{m_e}
  =9\times10^{18}\,.
\ee
We thus find that when entering the recombination regime at around $T\sim0.3\,\mathrm{eV}$, the electron temperature deviates only by $(T-T_e)/T=1.6\times10^{-7}$ from the photon temperature. The electrons, in other words, do not in any significant way affect the photon temperature and the photons thus follow an equilibrium distribution into the recombination regime, explaining the observed black-body nature of the CMB. This argument can be made  more quantitative by considering the Boltzmann equation for the photons which, in this regime, is known as the Kompaneets equation \cite{kompaneets}: 
\be
  \label{komp}
  \left(\partial_t-H \omega\partial_\omega\right)g(\omega,t)=\frac{8\pi\alpha_\mathrm{em}^2}{3m_e^3}n_e\frac{1}{\omega^2}\partial_\omega\left[\omega^4\left(T_e\partial_\omega g(\omega,t)+\left[1+g(\omega,t)\right]g(\omega,t)\right)\right]\,,
\ee
where $g$ and $\omega$ are the photon distribution function and energy, respectively. One may now \emph{define} the photon temperature by the relation
\be
  g(1+g)=-T\partial_\omega g\,,
\ee
which is automatically satisfied for a thermal distribution. With this definition, one realizes that for $T=T_e$ the collision term on the right hand side of \kref{komp} vanishes identically and the photons therefore just follow an equilibrium distribution.

\subsection{Neutralino dark matter}

One of the most well-motivated and extensively studied DM candidates is the neutralino. In most supersymmetric extensions to the SM, it appears as the lightest supersymmetric particle (LSP) and is given by   
 a linear combination of the superpartners of the gauge and Higgs fields,
\begin{equation}
  \label{neut}
  \chi\equiv\tilde\chi^0_1= N_{11}\tilde B+N_{12}\tilde W^3 +N_{13}\tilde H_1^0+N_{14}\tilde H_2^0\,.
\end{equation}
Here, in order to illustrate our formalism and to describe the kinetic decoupling process for a typical WIMP dark matter candidate, we consider the simple case of a Bino-like neutralino, $\chi\sim\tilde B$. From the WMAP relic density constraint, Binos have a mass of a few hundred GeV or less.

To lowest order in the fermion energy $\omega$, the scattering amplitude of Binos with SM fermions is given by \cite{Hofmann:2001bi}:
\be
  \left|\mathcal{M}\right|^2_{t=0} = 32g_Y^4\left(Y_d^4+ Y_s^4\right)\left(\frac{M_{\tilde B}}{M^2_{\tilde L}-M^2_{\tilde B}}\right)^2\omega^2\,.
\ee
From \kref{TCDMmain}, we thus find that the neutralino temperature evolves with the background temperature as
\be
  \label{TLSP}
  T_\chi/T= \left\{1-\frac{z^{1/4}}{4} \exp[z]\, \Gamma\left[-1/4,z\right]\right\}_{z=\frac{a}{4}\left(T/M_{\tilde B}\right)^{4}}\,,
\ee
where
\bea
  a &\stackrel{(\ref{adef})}{=}& \frac{31}{42}\sqrt{\frac{5\pi^3}{g_\mathrm{eff}}}\frac{M_\mathrm{Pl}}{M_{\tilde B}}g_Y^4\sum g_\mathrm{SM}(Y_d^4+Y_s^4)\left(\frac{M^2_{\tilde B}}{M^2_{\tilde L}-M^2_{\tilde B}}\right)^2\\
    &=&2.7\times10^{16}\,\left(\frac{M^2_{\tilde B}}{M^2_{\tilde L}-M^2_{\tilde B}}\right)^2\left(\frac{M_{\tilde B}}{100\,\mathrm{GeV}}\right)^{-1}\,.
\eea
This corresponds to a decoupling temperature of
\be
 \label{Binokd}
 T_{\mathrm{kd}} \stackrel{(\ref{tdecmain})}{=}8.9 \times \left(\frac{M^2_{\tilde L}}{M^2_{\tilde B}}-1\right)^{1/2}\left(\frac{M_{\tilde B}}{100\,\mathrm{GeV}}\right)^{5/4}\,\mathrm{MeV}\,.
\ee
When computing the sum over all contributions, we have assumed an identical mass for all relevant sleptons $\tilde L$ (i.e.~(anti-)selectrons and (anti-)sneutrinos).

\begin{figure}
    \begin{center}
       \psfrag{x}[t][][1]{$T$ [MeV]}
       \psfrag{y}[b][][1]{$T_\chi/T$}
     \includegraphics[width=0.7\textwidth]{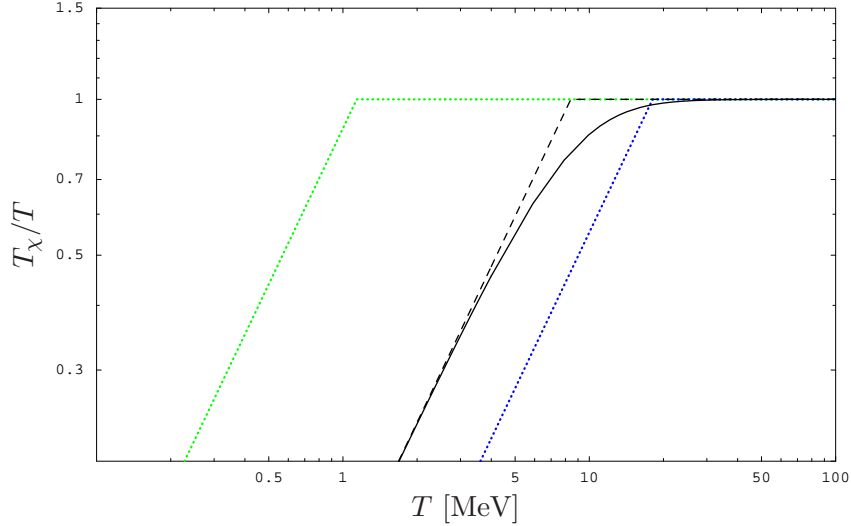}
    \end{center}
     \caption{The LSP temperature $T_\chi$ as a function of the background temperature $T$ in the early universe is shown as a solid line, for the case of a Bino with $M_{\tilde B}=100~$GeV and $M_{\tilde L}=150~$GeV. The rather abrupt transition to a regime with $T_\chi\propto T^2$, where the LSP is completely decoupled from the thermal bath, makes it straightforward to identify the temperature $T_\mathrm{kd}$ of thermal decoupling as given in Eq.~\kref{Binokd}; here, it is indicated by a dashed line. For comparison, the (green and blue) dotted lines show the naive estimates for the evolution of $T_\chi$, assuming a sudden decoupling from the thermal bath at the time of last scattering or the relaxation time, respectively (see also \cite{Hofmann:2001bi}). }
   \label{fig_LSP1}
\end{figure}

In Fig.~\ref{fig_LSP1}, we show the evolution of the CDM temperature for a typical Bino DM candidate during the process of kinetic decoupling. For reference, we also indicate the naive estimates for the decoupling temperature that were reviewed in Section \ref{sec_dec}. At temperatures only slightly greater than $T_{\mathrm{kd}}$, the LSP is still very tightly bound to the background temperature $T$ (for reference, see also Fig.~\ref{fig_phaseplot}). We therefore expect our result to be highly independent of the details of the QCD phase transition at around $T_\mathrm{QCD}\sim150~$MeV. A second important observation is the rather sharp turnover from  $T_\chi\approx T$ to a regime with $T_\chi\propto T^2$, where the LSP has completely decoupled from the thermal bath and simply cools down due to the Hubble expansion. Finally, to take up our previous discussion on the various scales that are involved in the decoupling process, we can clearly see that the relevant quantity  is not the time of last scattering, which happens long after the LSP has already decoupled. The relaxation time $\tau_\mathrm{rel}$, on the other hand, gives as expected a much better estimate for the time of decoupling. The reason that it \emph{underestimates} the actual decoupling time is to be found in the fact that close to decoupling only small deviations from a thermal distribution need to be restored -- while the derivation of the relaxation time $\tau_r$ implicitly assumed that thermal equilibrium had to be restored from a highly non-thermal distribution.

Obviously, our formalism can in principle easily be extended beyond the simple Bino case that we have presented here and applied to an arbitrary neutralino candidate. In particular, it would be interesting to perform a scan over, e.g., the MSSM parameter space to better understand the expected spread of the decoupling temperature.  We note that a first step into this direction has been performed in Ref.~\cite{Profumo:2006bv}, based on the estimate that the relaxation time as defined before is a good approximation for the decoupling time; in the case of the Bino, this would correspond to a kinetic decoupling temperature of \cite{Hofmann:2001bi}
\be
  T_\mathrm{kd}\sim \left[1.2\times10^{-2}\frac{M_\mathrm{Pl}}{M_\chi\left(M^2_{\tilde L}-M^2_{\tilde B}\right)^2}\right]^{-1/4}\,,
\ee
which we have, for comparison, also included in Fig.~\ref{fig_LSP1}. We would like to stress that it would be warranting to  re-evaluate the results of Ref.~\cite{Profumo:2006bv} in light of the method that we have presented here; in particular, since we are now able to actually follow the details of the kinetic decoupling process, the underlying uncertainty in the determination of the cutoff-scale is greatly reduced. Also, the method can be applied to regions of the parameter space where a simple scaling of the cross section with energy, like $\sigma\sim const.$ or $\sigma\propto E_l^2$, is not valid anymore (see \ref{app_LKPscatter}).

\subsection{Kaluza-Klein dark matter}
\label{subsec_LKP}

Models with universal extra dimensions (UED) \cite{Appelquist:2000nn}, where all standard model (SM) fields are allowed to propagate in a higher-dimensional bulk, have received a great deal of attention since it was realized that they naturally give rise to a new class of dark matter candidates \cite{Cheng:2002iz,Servant:2002aq}: the higher-dimensional extra degrees of freedom appear in the low-energy effective 4D theory as towers of new, heavy states, the lightest of which -- similar to the case of $R$-parity in supersymmetry -- is stable due to an internal $Z_2$ symmetry (``KK-parity'') that appears as a remnant of the higher-dimensional translational invariance; thermally produced in the early universe, the lightest Kaluza-Klein particle (LKP) acquires  the right WMAP relic density for a compactification scale of about $R^{-1}\sim1\,$TeV \cite{Kong:2005hn,Burnell:2005hm,Kakizaki:2006dz}.

At tree-level, all Kaluza-Klein modes are essentially degenerate in mass, $M\approx R^{-1}$, so radiative corrections become important for the determination of the actual mass hierarchy between these states and, in particular, for the determination of the LKP. Unfortunately, the radiative mass shifts depend in principle on the (unknown) physics at the cutoff scale $\Lambda$ and should in that sense be treated as free -- though small -- parameters. However, one usually follows the (self-consistent) assumption that the corresponding contributions at that scale are negligible \cite{Cheng:2002iz}. In this minimal UED (mUED) model, all masses are determined by only two parameters, $R$ and $\Lambda$, and the LKP turns out to be the $B^{(1)}$, the first KK excitation of the weak hypercharge gauge boson. 

\begin{figure}[t]
  \hfill
  \begin{minipage}[t]{.45\textwidth}
    \begin{center}  
       \psfrag{x}[t][][1]{$1/R$}
       \psfrag{y}[b][][1]{$T_\mathrm{kd}$ [MeV]}
     \includegraphics[width=\textwidth]{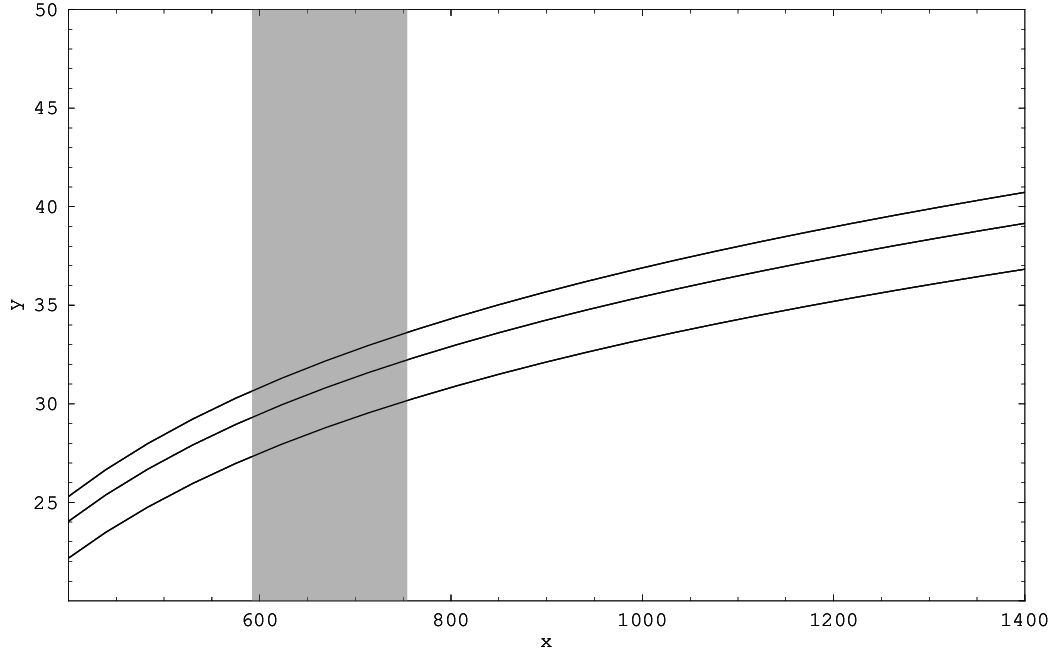}
     \end{center}
  \end{minipage}
  \hfill
  \begin{minipage}[t]{.45\textwidth}
     \begin{center}  
       \psfrag{x}[t][][1]{$M$}
       \psfrag{y}[b][][1]{$T_\mathrm{kd}$ [MeV]}
     \includegraphics[width=\textwidth]{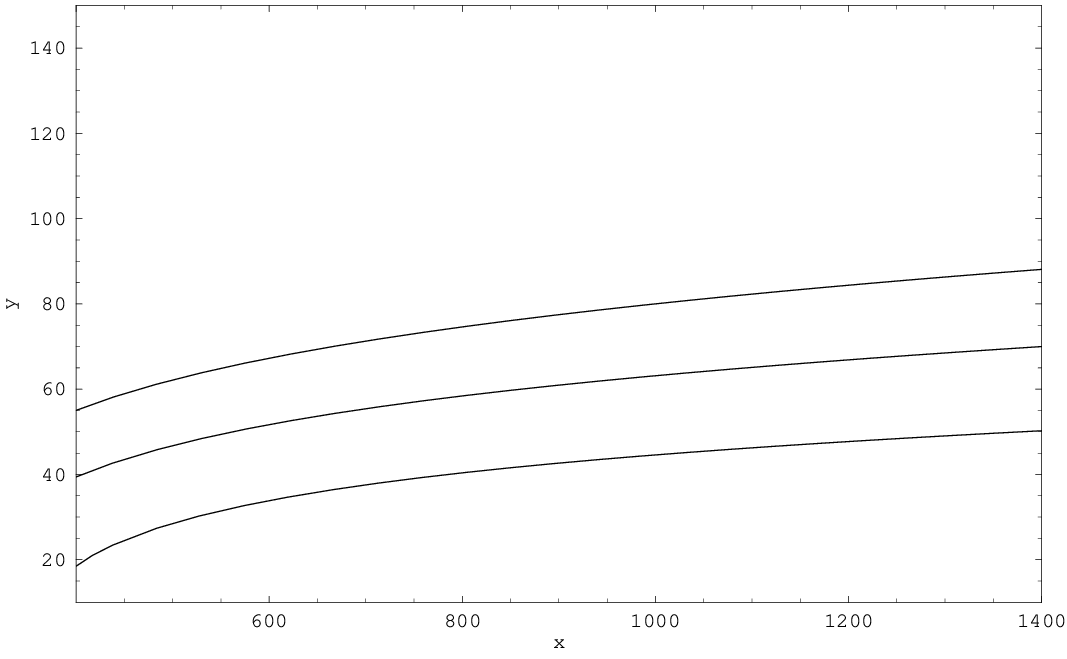}
     \end{center}
  \end{minipage}
  \hfill
  \caption{The left part of this figure shows the kinetic decoupling temperature $T_\mathrm{kd}$ for the LKP in the mUED model. From top to bottom, the cutoff scale is set to $\Lambda R=40,\,30,\,20$ and the grey band shows the region  consistent with the $2\sigma$ WMAP relic density constraint for a Higgs mass $m_h\lesssim150\,$GeV (for higher Higgs masses, the grey region broadens and shifts to the right, up to $R^{-1}\lesssim1.3\,$TeV \cite{Kakizaki:2006dz}). In the right panel, we leave the mUED setup and treat the mass splitting between the $B^{(1)}$ and the KK leptons as a single, free parameter $\delta$; from bottom to top, the curves correspond to $\delta=0.01,\,0.05,\,0.2$.  }
\label{fig_LKP}
\end{figure}

In \ref{app_LKPscatter}, we calculate the scattering amplitude for the LKP with SM fermions; to lowest order in the fermion energy $\omega$, it is given by :
\bea
  \left|\mathcal{M}\right|^2_{t=0}&=&\frac{16}{3}g_Y^4\left\{Y_d^4\left(\frac{M_{B^{(1)}}}{M^2_{f_d^{(1)}}-M^2_{B^{(1)}}}\right)^2 + Y_s^4\left(\frac{M_{B^{(1)}}}{M^2_{f_s^{(1)}}-M^2_{B^{(1)}}}\right)^2\right\}\omega^2\\
    &\approx& \frac{4}{3}g_Y^4Y_s^4\delta_s^{-2} \omega^2\,,
\eea
where in the last step we used the fact that the fermion singlet KK states receive much smaller radiative corrections to their masses than the KK doublet states and introduced
\be
  \delta_s\equiv (M_{f_s^{(1)}}-M_{B^{(1)}})/M_{B^{(1)}}\ll1\,.
\ee
We note that the amplitude takes the same form as for a Bino LSP. The evolution of the LKP temperature is thus described by Eq.~\kref{TLSP}, but now with
\bea
  \label{aLKP}
  a &=& \frac{31}{252}\sqrt{\frac{5\pi^3}{g_\mathrm{eff}}}\frac{M_\mathrm{Pl}}{M_{B^{(1)}}}g_Y^4\sum g_\mathrm{SM}\left\{Y_d^4\left(\frac{M^2_{B^{(1)}}}{M^2_{f_d^{(1)}}-M^2_{B^{(1)}}}\right)^2 + Y_s^4\left(\frac{M^2_{B^{(1)}}}{M^2_{f_s^{(1)}}-M^2_{B^{(1)}}}\right)^2\right\}\\
  &\approx& 10^{14}\,\delta_s^{-2}\left(\frac{M_{B^{(1)}}}{\mathrm{TeV}}\right)^{-1}\,.
\eea
The corresponding decoupling temperature is therefore given by
\be
 T_{\mathrm{kd}} \stackrel{(\ref{tdecmain})}{\approx}3\times10^2\, \delta_s^{1/2}\left(\frac{M_{B^{(1)}}}{\mathrm{TeV}}\right)^{5/4}\,\mathrm{MeV}\,.
\ee
The sum that appears above runs over all SM fermions that contribute to the scattering process; for temperatures around the kinetic decoupling, these are only electrons and the three neutrino species, as well as the corresponding antiparticles (the contribution from all other particles is strongly Boltzmann suppressed). The mass-splitting between the LKP and the corresponding KK states is then given by $\delta\sim10^{-2}$. In Figure \ref{fig_LKP}, we take the exact mUED values for $\delta$ and show $T_{\mathrm{kd}}$ as a function of $R$ and $\Lambda$, using the full expression \kref{aLKP}. We find that the LKP decoupling temperature falls into a very narrow range,
\be
 27\,\mathrm{MeV} \lesssim T_{\mathrm{kd}}\lesssim 34\,\mathrm{MeV}\,,
\ee
depending on the parameters of the model. The low value of $T_{\mathrm{kd}}$, despite a TeV scale LKP mass, justifies the above statement about the number of relativistic SM degrees of freedom close to decoupling. In particular, we would like to stress that the kinetic decoupling of the LKP -- just as for the Bino -- happens sufficiently late so as not to be influenced by the details of the QCD phase transition at around $T_\mathrm{QCD}\sim150$~MeV.

As a next step, we relax the assumptions of the mUED setup and treat the LKP mass and the mass splitting as independent parameters that may be varied about the mUED values. As can be seen from the right panel of Fig.~\ref{fig_LKP}, one may in this way encounter higher decoupling temperatures -- but without invoking extreme fine-tuning, one will still stay below $T_\mathrm{kd}\lesssim100\,$MeV. Note, however, that once one leaves the mUED setup, one may in principle also consider other LKP DM candidates than the $B^{(1)}$. The only reasonable choices in this case are the KK excitations of the $Z$ and Higgs boson, respectively (direct detection experiments have already ruled out the KK neutrino as a DM candidate \cite{Servant:2002hb}). For the $Z^{(1)}$, we expect a very similar phenomenology as for the $B^{(1)}$, with the $U(1)$ coupling $g_Y$ replaced by the (stronger) $SU(2)$ coupling $g$. From \kref{Tkd_OM}, this results naively in a smaller decoupling temperature. However, we want to keep the annihilation cross section $\left<\sigma v\right>\propto \alpha^2/M^2$ roughly constant in order to satisfy the relic density contraint. The expected net dependence on the coupling is thus roughly given by
\be
  T_\mathrm{kd}\propto\alpha^{-1/2}M^{5/4}\propto\alpha^{3/4}\,.
\ee
For a $Z^{(1)}$ LKP, this corresponds to an \emph{increase} of $T_\mathrm{kd}$ by a factor of about $(g/g_Y)^{3/2}\sim$2.5 as compared to the case of the $B^{(1)}$. The situation of a KK Higgs boson as DM candidate is more complicated, already because the annihilation rate does no longer scale in the simple way indicated above, and we leave it therefore open for future studies.

\section{From kinetic decoupling to damping scales in the power spectrum}
\label{sec_clumps}

After having provided a detailed description of the kinetic decoupling process in the preceeding part of this article, we would now like to briefly discuss in the next two Sections, on a rather qualitative level, the further evolution of the WIMP population in the expanding universe.

To start with,
and in order to get a qualitative intuition for the physical processes leading to kinetic damping of CDM perturbations, let us use the hydrodynamical description for discussing the evolution of these perturbations. 
For temperatures $T>T_{\rm kd}$, i.e.~before kinetic decoupling, small-scale perturbations in the CDM fluid are damped due to the tight
coupling between the CDM and radiation fluid. Note that this includes temperatures $T<M_\chi$, when the CDM fluid
is already non-relativistic. Around kinetic decoupling, $T\sim T_{\rm kd}$, the CDM fluid starts to deviate from an ideal fluid, i.e.~the
coupling to the radiation component is not strong enough to maintain local thermal equilibrium for the CDM any longer. However, the residual coupling between
these two fluids generates some friction between them. The main viscous processes causing this friction are bulk- and shear-viscosity,
while heat conduction is subdominant.  
Hence, perturbations of the CDM energy momentum tensor behave like sound waves propagating through an opaque medium 
(the heat bath) \cite{Weinberg:1971mx}. Their dispersion relation receives in-medium modifications, resulting in an imaginary frequency that is generated by the viscous coupling
between the CDM perturbations and the heat bath. 

The strength of the viscous couplings can not be calculated within the hydrodynamical framework. The most convenient strategy for
calculating the viscosity coefficients is to calculate the entropy generation in the CDM-radiation mixture within kinetic theory
and equate it to the corresponding expression derived in the hydrodynamical framework. This allows to simply read off
the viscosity coefficients by comparing the relevant tensors in both approaches that describe, respectively, the same type of entropy generation.

The effect of the viscous coupling is to damp perturbations in the CDM energy momentum tensor; in fact, they generate an
exponential cut-off in the perturbation spectrum at kinetic decoupling. This spectrum then serves as initial condition for
the subsequent evolution at temperatures $T<T_{\rm kd}$ -- the so-called free streaming regime, where a further damping in the perturbation spectrum take place. The free streaming of CDM particles leaves a finger-print  similar to collisional
damping in the spectrum, i.e. a characteristic exponential damping term.
Free streaming is therefore often also referred to as collisionless damping. 
For WIMPs, the characteristic damping mass is set by
\be
\frac{M_{\rm fs}}{10^{-6}M_\odot} 
\approx 
\left[
\frac{1+{\rm ln}\left(T_{\rm kd}/30\;{\rm MeV}\right)/19.2}
{\left(M_\chi/100\; {\rm GeV}\right)^{1/2} \left(T_{\rm kd}/30\;{\rm MeV}\right)^{1/2}}
\right]^3
\; .
\ee 
The characteristic damping scale for free streaming is time dependent, however, it becomes
constant to a very good approximation sufficiently late after matter radiation equality.

The authors of \cite{Green:2003un,Green:2005fa} calculated the envelope of the CDM perturbation spectrum. In \cite{Loeb:2005pm}
 it was later shown numerically
that acoustic oscillations have to be taken into account at kinetic decoupling as they give rise to even stronger
damping; the order of magnitude of the cut-off scale, however, is correctly given in \cite{Green:2003un,Green:2005fa}. 
Most recently, the author of \cite{Bertschinger:2006nq} succeeded in including the acoustic oscillations before and during kinetic
decoupling, the viscous coupling during kinetic decoupling and free streaming in a consistent
kinetic treatment. All approaches agree in that the dominant suppression of small-scale power is given by
collisionless damping during free streaming, with kinetic decoupling providing consistent initial
conditions. 

During radiation domination and on sub-Hubble scales, the transfer function for matter density perturbations grows logarithmically,
while after matter radiation equality matter density perturbations increase linearly (for a vanishing baryon fraction) with the scale factor on sub-Hubble
scales. In both eras, during radiation and during matter domination, the spectrum of perturbations depends logarithmically on
the comoving wavenumber that characterizes the corresponding mode. Hence, modes characterized by wavenumbers close to the cut-off
in the power spectrum typically enter the nonlinear regime first. This is the basic picture of hierarchical structure formation:
small scales collapse first to nonlinear structures, merging only later to larger structures. For WIMPs, this typically happens at a redshift  between 40 and 80,
depending on the interaction theory underlying the WIMP candidate and the details of the primordial power spectrum.
All matter density perturbations that grow nonlinear are characterized by masses above $M_{\rm fs}$. 

Whether these first purely gravitationally bound WIMP structures can survive until today is under debate. 
Their mean density contrast is comparable to that of the galactic disk and an order of magnitude above that of the halo
in the solar neighborhood. One has to keep in mind, however, that the very first perturbations entering the nonlinear regime are, rather, 
rare fluctuations with mass scales $M_\mathrm{rf}>M_\mathrm{fs}$. Those rare fluctuations can easily reach density contrasts which are an order of magnitude higher.
High resolution simulations \cite{Diemand:2005vz} show that structures with physical characteristics set by the cut-off in the linear
power spectrum may survive merger processes of the first halos. The next critical phase starts with the formation
of stars. Encounters with stars could disrupt the first CDM halos. So far, various studies \cite{Green:2006hh,Goerdt:2006hp} show that the first halos
would loose mass during an encounter, but their core should remain intact.

\section{Impacts for the indirect detection of dark matter}
\label{sec_inddet}

Indirect detection methods for DM rely on the fact that most particle DM candidates can pair-annihilate into standard model particles. The annihilation products could then potentially be spotted in cosmic rays of various kinds, with an expected flux contribution that is proportional to the DM density squared. From  hierarchical structure formation,  typical Milky-Way sized Halos are expected to contain a plethora of DM substructures rather than a homogeneous DM distribution.
Focussing in the following on gamma rays, the consequences for indirect DM detection are two-fold and, as we will see, depend sensitively on an accurate determination of the cutoff scale in the power-spectrum as we have performed it in the preceeding part of this work. 

At this point, let us briefly introduce some necessary notation and denote by $\rho_\mathrm{sub}(M_\mathrm{sub},\mathbf{r}_\mathrm{sub})$ the
 density profile inside a typical dark matter subhalo (or ``clump'') of mass
$M_\mathrm{sub}$ (as opposed to the smooth, i.e., average DM density $\rho(\mathbf{r})$ in the Milky Way), and by $n_\mathrm{sub}(M_\mathrm{sub},\mathbf{r})$ the
number density of such clumps at a position $\mathbf{r}$ in the
halo. For subhalos that cannot be resolved separately, the average annihilation flux originating from $\mathbf{r}$ is then proportional to $\rho^2_\mathrm{eff}(\mathbf{r})$, rather than $\rho^2(\mathbf{r})$, where 
\bea
  \rho^2_\mathrm{eff}(\mathbf{r}) &\equiv& \int\mathrm{d}M_\mathrm{sub}\int\mathrm{d}^3\mathbf{r}_\mathrm{sub}\,n_\mathrm{sub}(M_\mathrm{sub},\mathbf{r}-\mathbf{r}_\mathrm{sub})(\rho_\mathrm{sub}(M_\mathrm{sub},\mathbf{r}_\mathrm{sub}))^2\label{clump2}\\
  &\simeq&\rho_0\int\mathrm{d}M_\mathrm{sub}\,M_\mathrm{sub}\delta(M_\mathrm{sub})n_\mathrm{sub}(M_\mathrm{sub},\mathbf{r})\label{smallclump}\,.
\eea
The dimensionless quantity
\be
  \label{deltadef}
  \delta\equiv\frac{1}{\rho_0}\frac{\int\mathrm{d}^3\mathbf{r}_\mathrm{sub}\left(\rho_\mathrm{sub}(M_\mathrm{sub},\mathbf{r}_\mathrm{sub})\right)^2}{\int\mathrm{d}^3\mathbf{r}_\mathrm{sub}\,\rho_\mathrm{sub}(M_\mathrm{sub},\mathbf{r}_\mathrm{sub})}
\ee
that appears here is a measure of the effective density contrast between an average
dark matter clump and the local halo density $\rho_0$. 
The last step in
\kref{smallclump} is valid if $n_\mathrm{sub}$
does not change very much on scales of the order of the size of a
clump, i.e.~for the whole
 region of integration one has
$n_\mathrm{sub}(M_\mathrm{sub},\mathbf{r}-\mathbf{r}_\mathrm{sub})\approx
n_\mathrm{sub}(M_\mathrm{sub},\mathbf{r})$.

Equipped with the above notations, we can now turn our discussion to the \emph{total gamma-ray flux} from DM annihilations which is potentially strongly enhanced w.r.t.~the case of a smooth halo distribution. To see this, note that the number density of subhalos scales with their mass as $n_\mathrm{sub}\propto M_\mathrm{sub}^{-2}$ -- a behaviour that is expected from the theory of cosmological perturbations  and has recently numerically been confirmed down to a mass scale of $4\times10^6\,M_\odot$, corresponding to the resolution limit of state-of-the-art $N$-body simulations \cite{Diemand:2006ik}. To a first approximation, the density contrast $\delta$, on the other hand, scales only weakly with $M_\mathrm{sub}$; from \kref{smallclump}, this means that the luminosity of individual subhalos is proportional to $M_\mathrm{sub}$, leading to an equal contribution to the total subhalo-induced gamma-ray flux from each decade in subhalo mass -- which is precisely the behaviour that has numerically been found in \cite{Diemand:2006ik}. As we have stressed before, it is therefore crucial to have an exact and reliable method to determine the cutoff in the spectrum below which there are no contributions to the annihilation flux. Extrapolating the results of the above-mentioned study down to the typical cutoff mass scales that we found earlier, $M_\mathrm{fs}\sim 10^{-6}\,M_\odot$, we expect a factor of almost 4 for the increase of the expected annihilation signal as compared to that from a smooth halo distribution. Taking into account the effects of sub-subhalos may further enhance the annihilation flux significantly; on the other hand, reduced survival probabilities for very small clumps, as discussed in the previous Section, would lead to slightly less optimistic predictions. Each of these issues is far from being settled at the moment of writing and warrants more dedicated future studies on their own.

The other -- perhaps even more important -- effect of a large fraction of substructures in the DM halo distribution is the possibility of observing single DM clumps as gamma-ray point sources, maybe already in the near future with upcoming experiments like GLAST \cite{glast}. Comparing the projected all-sky GLAST sensitivity map for point-sources of DM annihilations \cite{GLAST_map} with the simulated gamma-ray sky of a Milky Way-like halo \cite{Diemand:2006ik}, unfortunately seems to suggest that, in order to actually resolve the brightest substructures, point-source sensitivities about two to three orders of magnitude better than the GLAST performance are needed. One has to keep in mind, however, that the flux estimates obtained in  \cite{Diemand:2006ik} are extremely conservative in the sense that no substructures smaller than $\sim10^4\,M_\odot$ are resolved. This means, in particular, that the density profile \emph{within} the clumps is essentially cut off at that scale; actual physical DM subhalos will exhibit much higher central densities and thus considerably enhanced fluxes (see also \cite{Diemand:2006ik} for similar conclusions).

In order to quantify these last remarks, we adopt for the subhalo profiles the usual $(\alpha,\beta,\gamma)$ parametrization,
\be
  \label{prof}
\rho_\mathrm{sub}(\mathbf{r})=\rho_\mathrm{sub}(r_\mathrm{s})\left(\frac{r}{r_\mathrm{s}}\right)^{-\gamma}\left[1+\left(\frac{r}{r_\mathrm{s}}\right)^\alpha\right]^\frac{\gamma-\beta}{\alpha}\,,
\ee
and assume it to be valid down to a scale $r_\mathrm{cut}=\max\{r_\mathrm{kd},r_\mathrm{ann}\}$ (here, $r_\mathrm{kd}$ corresponds to the earlier determined cut in the power spectrum due to damping effects after the kinetic decoupling of WIMPs from the heat bath, and $r_\mathrm{ann}$ is set by the maximal possible DM density due to DM self-annihilations \cite{ber}). The scale radius $r_\mathrm{s}$ determines  the transition from large ($\rho_\mathrm{sub}\propto r^{-\beta}$) to small ($\rho_\mathrm{sub}\propto r^{-\gamma}$) radial distances; for the small subhalos we are interested in here, one expects rather small concentration parameters, $c\equiv r_\mathrm{vir}/r_\mathrm{s}<3$ \cite{die}, which means that the main contribution to the annihilation flux originates from the innermost regions of the subhalos and we can, for our purposes, well approximate the profile as $\rho_\mathrm{sub}\propto r^{-\gamma}$. Since the annihilation flux from a single subhalo is proportional to $\delta$, as defined in Eq.\kref{deltadef}, we don't expect a strong dependence of this flux on the value of $r_\mathrm{cut}$ for the case of mildly cuspy profiles ($\gamma=1$). For a more cuspy profile like $\gamma=1.5$, on the other hand, we expect the annihilation flux to scale logarithmically with $r_\mathrm{cut}$, thus leading to an enhancement of up to 30 w.r.t. to the flux considered in \cite{Diemand:2006ik}. In fact, at least at higher redshifts, one expects even steeper profiles for small subhalos, with $\gamma$ ranging from 1.5 to 2 \cite{die}, thus potentially further enhancing the resulting flux.

Of course, one has to keep in mind that neither the details of the innermost subhalo profiles nor the survival properties of these objects are sufficiently known at the present moment to make firm predictions. The numbers that we have presented above should therefore be considered as rather naive estimates, meant to illustrate the potential for GLAST -- and certainly for successive, next-generation gamma-ray telescopes -- to detect particularly bright subhalos as gamma-ray point sources. Note, in particular, that once GLAST has detected promising looking point sources, these could further be studied with air Cerenkov telescopes (ACTs) like CANGAROO~\cite{canga}, HESS~\cite{hess}, MAGIC~\cite{magic} or VERITAS~\cite{veritas}, taking advantage of the much better sensitivity of these telescopes as compared to satellite based experiments (up to three orders of magnitude at high energies -- see, e.g. \cite{DMreviews}).

\section{Conclusions}
\label{sec_conc}

The dominant damping effect for the evolution of linear CDM perturbations is given by collisionless damping due to the geodesic motion
of CDM particles. Even though collisional damping is subdominant, it has to be taken into account during the kinetic decoupling process, which provides consistent initial conditions for 
the free streaming regime. More importantly, collisional damping leads to a free streaming damping scale that depends
on the temperature at kinetic decoupling. It is thus really this temperature scale, $T_\mathrm{kd}$, that renders the position of the cut-off scale
in the linear power spectrum sensitive to the WIMP microphysics. 
Since the cut-off scale might have observable consequences on sub-pc scales, at least in principal, 
the kinetic decoupling history provides an exciting interface between the WIMP interaction theory and
the substructures that are probed by direct and indirect dark matter searches.   

We have presented, for the first time, a complete and analytic description of the kinetic decoupling process,
without assuming a specific WIMP candidate; the master equation for the WIMP temperature evolution
has been derived from first principles and accommodates any WIMP interaction theory. 
This could be achieved by separating slow and fast processes in the collision integral by utilizing a
Born-Oppenheimer-type of ansatz, which proved to be an excellent approximation scheme
for analyzing the WIMP radiation mixture -- allowing, in particular, for a full error management; in this article, we have determined the decoupling scale to a precision of
$p^2/M_\chi^{\; 2}\sim T_{\rm kd}/M_\chi\sim 10^{-5}$.

The calculations presented here support the (order-of-magnitude) relaxation time approximation employed for the case
of bino-like WIMPs before \cite{Hofmann:2001bi}. However, as we have already mentioned, we were able to significantly reduce the uncertainty in the determination of the cut-off scale.
 Let us stress again that this holds for \emph{arbitrary} WIMP candidates and elastic scattering amplitudes.
In particular, the method presented here can even be applied to regions of the parameter space 
where a simple scaling of the total cross-section with energy, e.g. $\sigma\sim \omega^0$ or
 $\sigma\sim \omega^2$ is no longer valid -- as, e.g., in the case of $s$-channel resonances for LKP-like WIMPs. In this context, we would like to stress once more that it is, in fact, important to take into account the appropriate microphysics; wrong assumptions about the particle nature of the DM candidate can lead to grossly wrong results in the determination of the cutoff scale.
 
 Turning to explicit models, we have performed a complete scan over the mUED parameter space and found a very narrow range
 for the kinetic decoupling temperature: $27\,\mathrm{MeV}\lesssim T_{\rm kd}\lesssim34\,\mathrm{MeV}$. A parameter scan over the MSSM using the methods presented here
 lies beyond the scope of this work, however, see \cite{Profumo:2006bv}  for a current state of the art approach. Another interesting direction for future studies would be to modify the framework that we have presented here for WIMPs, in such a way as to predict the cutoff scale even for DM candidates like axions from first principles.
 
 Finally, we have commented on indirect detection prospects of a boosted signal from orthodox locations
 of enhanced dark matter densities, e.g. the galactic center or dwarf galaxies,
 and, much more promising, from individual dark matter clumps as gamma ray point sources. It remains, in fact, an exciting possibility that such objects may be seen already with the next generation of gamma-ray telescopes.

\vspace*{1.5cm}
\noindent
{\it Noted added. ---}
Subsequently to the publication of this article \cite{Bringmann:2006mu}, the limitation to 
relativistic scattering partners was overcome \cite{Bringmann:2009vf}. The collision term stated in Eq.~(B.20) 
then becomes (initial and final spin states are now being {\it summed} over)\footnote{
If one includes $g_\chi$ in the definition of $n_\chi$ and $T_\chi$, e.g.~$3M_\chi\,T_\chi n_\chi\equiv g_\chi\int \frac{d^3p}{(2\pi)^3}\,\mathbf{p}^2 f(\mathbf{p})$, this collision term should be divided by a further
factor of $g_\chi$.
}
\bea
  C(T) &=& \frac{1}{96\pi^3TM_\chi^2}\int d\omega\,k^4g^\pm(\omega)
  \left(1\mp g^\pm\right)
  \mathop{\hspace{-13.5ex}\left|\mathcal{M}\right|^2_{t=0}}_{\hspace{4ex}s=M_\chi^2+2M_\chi\omega+m_\mathrm{SM}^2}\nonumber\\
    &&\times \Big[M_\chi T\Delta_\mathbf{p} +\mathbf{p}\cdot\nabla_\mathbf{p}+3\Big]\,f(\mathbf{p})\\
    &\equiv&2M_\chi\gamma(T)\Big[M_\chi T\Delta_\mathbf{p} +\mathbf{p}\cdot\nabla_\mathbf{p}+3\Big] \,f(\mathbf{p}) \label{Cfullfinal}\,.
\eea
The above expression is only valid, to lowest order in the expansion parameter $T_\chi/M_\chi$, if $\left|\mathcal{M}\right|^2$ is Taylor expandable around $t=0$ (in the sense that  
$\left|\mathcal{M}\right|^2-\left|\mathcal{M}\right|^2_{t=0}\lesssim T^2_\chi/M^2_\chi \left|\mathcal{M}\right|^2_{t=0}$). If this assumption is relaxed, the collision term still takes the above form, but with the replacement \cite{Gondolo:2012vh,Kasahara}
\be
 \mathop{\hspace{-13.5ex}\left|\mathcal{M}\right|^2_{t=0}}_{\hspace{4ex}s=M_\chi^2+2M_\chi\omega+m_\mathrm{SM}^2}
 \longrightarrow\qquad
 \frac{1}{8k^4}\int_{-4k^2}^0\left|\mathcal{M}\right|^2(-t)dt\,, 
\ee
which describes an effective average over $t$.

Finally, a simplification in the original treatment \cite{Bringmann:2006mu} was to assume a constant effective number of heat bath degrees of freedom, which was later generalized \cite{Bringmann:2009vf,vandenAarssen:2012ag}. Assuming no non-standard entropy production, in particular, it is advantageous to consider the dimensionless parameters 
$x\equiv {M_\chi}/{T}$ and $y\equiv{M_\chi T_\chi}{s^{-2/3}}$,
because the main process equation -- Eqs.~(3) and (A.12) in \cite{Bringmann:2006mu} -- then takes a particularly simple form \cite{vandenAarssen:2012ag}:
\be
\frac{x}{y}\frac{dy}{dx}=-\left(1-\frac{x}{3}\frac{1}{g_*}\frac{d g_*}{dx}\right) \frac{\gamma(T)}{H}\left(1-\frac{y_\mathrm{eq}}{y}\right)
\,,
\ee
where $g_*$ is the effective number of entropy degrees of freedom of the heat bath.
This equation, with the collision term stated above, is valid in full generality (as long as the co-moving DM number density does not change during or after kinetic decoupling, like
for example in the presence of the Sommerfeld effect \cite{vandenAarssen:2012ag}). The kinetic decoupling temperature, in analogy with Eq.~(5), is then obtained as
\be
x_\mathrm{kd}=\frac{M_\chi}{T_\mathrm{kd}}=\left.y\right|_{x\to\infty}\left.\frac{s^{2/3}}{T^2}\right|_{T=T_\mathrm{kd}}\,.
\ee
Note that $y(x)$ typically converges very quickly to a constant value for $x>x_\mathrm{kd}$, such that this definition is both unique and rather independent of the assumed cosmological evolution
(in contrast to what is indicated in \cite{Visinelli:2015eka}), c.f.~Fig.~1 in \cite{Bringmann:2009vf}.

\vfill
\subsection*{Acknowledgements}
Research at Perimeter Institute for Theoretical Physics is supported in part by the Government of Canada through NSERC and by the Province of Ontario through MRI.

\newpage
\appendix

\section{Boltzmann equation and temperature evolution in a Friedmann-Robertson-Walker spacetime}
\label{app_kd}

The Boltzmann equation is one of the fundamental equations of non-equilibrium statistical mechanics and describes the evolution of the phase space distribution function $f(\mathbf{x},\mathbf{p},t)$ for a given particle species $\chi$. In an abstract form, it is usually written as
\be
  \label{boltzmann}
  \hat L[f]=C[f]\,.
\ee
Here, the Liouville operator $\hat L$ is the covariant generalization of the convective derivative familiar from hydrodynamics, or -- in more technical terms -- the variation with respect to an affine parameter $\lambda$ along a geodesic:
\be
  \hat L[f]= \frac{df}{d\lambda}=\frac{dx^i}{d\lambda}\frac{\partial f}{\partial x^i}+\frac{dp^i}{d\lambda}\frac{\partial f}{\partial p^i}=p^i\frac{\partial f}{\partial x^i}-\Gamma^i_{\rho\sigma}\,p^\rho p^\sigma\frac{\partial f}{\partial p^i}\,,
  \label{totdiff}
\ee
where in the last step we have chosen $\lambda=\tau$ (i.e.~the eigentime).
Note that the sum is only over spatial momenta $p^i$ since we consider on-shell particles and hence $p^0$ is not an independent variable of $f$; likewise, we only sum over spatial
coordinates $x^i$ because we consider a Hamiltonian system where $f$ cannot have an explicit time-dependence. In a (flat) Friedmann-Robertson-Walker (FRW) spacetime, 
\be
  ds^2=dt^2-a^2(t)d\mathbf{x}^2,
\ee
the homogeneity of space implies $f=f(\left|\mathbf{p}\right|,t)$ and Eq.~(\ref{totdiff})
becomes
\bea
  \hat L[f]&=&  -2p^0H\mathbf{p}\cdot \left.\nabla_\mathbf{p}f(\mathbf{p})\right|_{\mathbf{p}=\mathbf{p}(t)}\\
     &=& p^0\left(\partial_t-H\mathbf{p}\cdot\left.\nabla_\mathbf{p}\right)f(t,\mathbf{p})\right|_{\mathbf{p}=\mathbf{p}(t)}\,,
\eea
where we have introduced $H\equiv\dot a/a$. In the second step, we have adopted the standard convention of treating $f$ explicitly as a function of two $(t,\mathbf{p})$ rather than only one $(\mathbf{p})$ variable, using that {\it to leading order} the time-dependence
of $\mathbf{p}$ derives exclusively from the scalefactor $a$, i.e.~$\mathbf{p}(t)=\bar{\mathbf{p}}/a(t)$. Writing $f$ as a function of the co-moving momenta $\bar{\mathbf{p}}$ instead, $f=f(t,\bar{\mathbf{p}})$, the Liouville operator would
 simply become $\hat L=p^0\,\partial_t$. 

The collision term $C[f]$ on the right hand side of the Boltzmann equation takes account of all interactions that $\chi$ may experience. In \ref{app_collint}, we determine its exact analytical form for a non-relativistic particle that is in contact with a heat-bath of relativistic particles $F$  through two-body scattering processes of the type $F\chi\rightarrow F\chi$. As explained in Section~\ref{sec_dec}, this is precisely the situation that we encounter after the chemical decoupling of $\chi$ in the early universe. To leading order in  $T/M_\chi$ and $\mathbf{p}^2/M^2_\chi$, where $T$ is the temperature of the heat bath, the collision term then reads
\be
  \label{cgeneral}
  C[f]=c M_\chi^2\left(\frac{T}{M_\chi}\right)^{n+4}\left\{M_\chi T\Delta_\mathbf{p}+A\mathbf{p}\cdot\nabla_\mathbf{p}+B\right\}f(\mathbf{p})\,.
\ee
In the above expression, $c$ is an overall constant that is given in Eq.~\kref{Cfinal}. Furthermore, we actually have $A=1$ and $B=3$, but in order to develop a physical understanding for the value of these constants, we will, for the time being, keep our discussion to the general form displayed in Eq.~\kref{cgeneral}.

Let us start by dividing the Boltzmann equation by $p^0$ and then integrate it over all momenta. On the left-hand side, we find
\be
  \int \frac{d^3p}{(2\pi)^3}\left\{\partial_t-H\mathbf{p}\cdot\nabla_\mathbf{p}\right\}f(\mathbf{p})=a^{-3}\partial_t\left(a^3n_\chi\right)\,,
\ee
where $n_\chi$ is the number density of $\chi$. Performing the same integral over \kref{cgeneral} gives
\be
 \int \frac{d^3p}{(2\pi)^3}\, C[f]/p^0= c M_\chi\left(\frac{T}{M_\chi}\right)^{n+4}(B-3A)\,n_\chi\,.
\ee
Of course, scattering processes should not change the comoving number density $a^3n_\chi$ -- which, from the above expressions, is only true for $B=3A$.

Inspired by the situation for a thermal distribution, let us now \emph{define} the ``temperature'' $T_\chi$ for an arbitrary distribution $f$ of non-relativistic particles as
\be
  \int \frac{d^3p}{(2\pi)^3}\,\mathbf{p}^2 f(\mathbf{p})\equiv 3M_\chi\,T_\chi n_\chi\,.
\ee
With this definition, we have $T_\chi\approx T$  for a situation close to thermal equilibrium and $\epsilon\equiv(T-T_\chi)/T$ can therefore be considered as a parameter that characterizes the deviation of $f$ from a thermal distribution. To determine the evolution of $T_\chi$, we consider the next moment of the Boltzmann equation:
\be
  \int \frac{d^3p}{(2\pi)^3}\, \frac{\mathbf{p}^2}{p^0}\,\hat L[f]=\int \frac{d^3p}{(2\pi)^3}\, \frac{\mathbf{p}^2}{p^0}\,C[f]\,.
\ee
Performing the integrals, and using $B=3A$, we arrive at the following process equation:
\be
\label{process}
  T_{\chi}'-\left[2+a A\left(\frac{T}{M_\chi}\right)^{n+2}\right]\frac{T_\chi}{T}=-a\left(\frac{T}{M_\chi}\right)^{n+2},
\ee
with
\be
  a\equiv\left(\frac{45}{\pi^3 g_\mathrm{eff}}\right)^{1/2}\frac{M_\mathrm{Pl}}{M_\chi}\,c\,.
\ee
Here, $T_{\chi}'\equiv dT_{\chi}/dT$ and we have assumed that the universe is radiation dominated, with $g_\mathrm{eff}$ effective degrees of freedom in relativistic particles. For $n\neq-2$, Eq.~\kref{process} can be solved by the ansatz
\be
  T_\chi=\lambda(T)\,T_\chi^\mathrm{(hom)}=\frac{\lambda(T)}{M_\chi}T^2\,\exp\left[\frac{aA}{n+2}\left(\frac{T}{M_\chi}\right)^{n+2}\right].
\ee
The general solution to the inhomogeneous differential equation (\ref{process}) then reads:
\be
  T_\chi=\frac{T}{A}\left\{1-\frac{z^{1/(n+2)}}{n+2} \exp[z]\, \Gamma\left[-(n+2)^{-1},z\right]\right\}_{z=\frac{aA}{n+2}\left(\frac{T}{M_\chi}\right)^{n+2}}+\lambda_0T_\chi^\mathrm{(hom)}\,.
\ee
In the limit of large and small $T$, respectively, the incomplete Gamma function behaves as
\bea
 \Gamma\left[-(n+2)^{-1},z\right]&\stackrel{T\rightarrow\infty}{\longrightarrow}& z^{-1/(n+2)} \exp[-z]/z\,,\\
 \Gamma\left[-(n+2)^{-1},z\right]&\stackrel{T\rightarrow0}{\longrightarrow}& \frac{n+2}{z^{1/(n+2)}}+\Gamma\left[-1/(n+2)\right]\,.
\eea
Only for $A=1$, as obtained in \kref{Cfinal}, one can thus recover the expected asymptotic behaviour (by choosing $\lambda_0=0$):
\bea
   T_\chi&\stackrel{T\rightarrow\infty}{\longrightarrow}& T\label{asymptotics1}\,,\\
   T_\chi&\stackrel{T\rightarrow0}{\longrightarrow}& \left(\frac{a}{n+2}\right)^{1/(n+2)}\Gamma\left[\frac{n+1}{n+2}\right]\,\frac{T^2}{M_\chi}\label{asymptotics2}\,.
\eea
At high temperatures, the particles are still kept in thermodynamic equilibrium, i.e.~$T_\chi=T$. The scaling $T_\chi\propto T^2$ at low temperatures, on the other hand, corresponds to the solution of the collision-less Boltzmann equation, i.e.~Eq.~(\ref{process}) for $a=0$. 

\begin{figure}
    \begin{center}
       \psfrag{x}[t][][1]{$\log_{10}x$}
       \psfrag{y}[b][][1]{$\log_{10}y$}
     \includegraphics[width=0.7\textwidth]{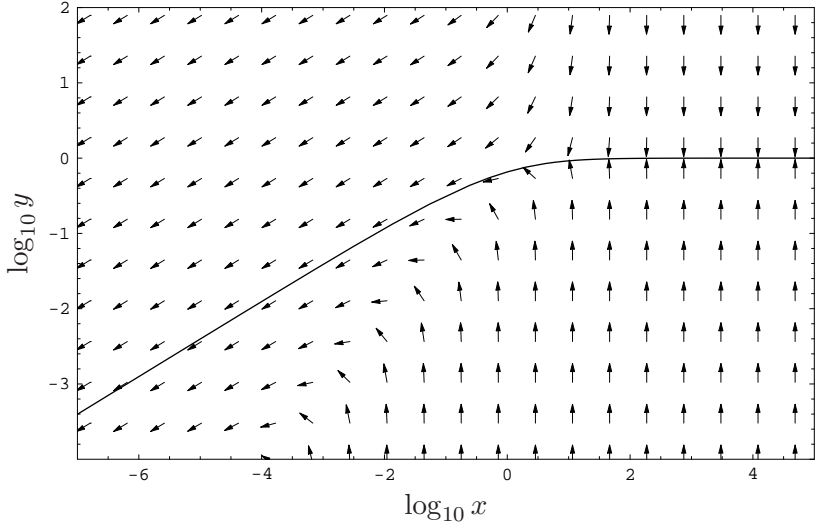}
    \end{center}
     \caption{Phaseplot for the temperature evolution, expressed in the dimensionless quantities $x\equiv\left|a\right|(T/M_\chi)^{n+2}$ and $y\equiv T_\chi/T$. Here, we show the case for $n=0$, but the same type of behaviour is obtained for any other (positive) value of $n$. As can be seen, any departure from thermal equilibrium ($T_\chi=T$) is restored almost immediately for $x\gtrsim10$, forcing $T_\chi$ to follow the solid line which represents our general solution \kref{TCDM}. For smaller temperatures, $\chi$ decouples from the heat bath; in this regime, its temperature is only influenced by the Hubble expansion and thus evolves as $T_\chi\propto T^2$.}
   \label{fig_phaseplot}
\end{figure}

To conclude our discussion of the process equation \kref{process}, let us rewrite it as
\be
  \label{process2}
  (n+2)\frac{dy}{dx}=\frac{y}{x}\pm(y-1)\,,
\ee
where we have introduced dimensionless variables
\be
   x\equiv\left|a\right|(T/M_\chi)^{n+2}\qquad\mathrm{and}\qquad y\equiv T_\chi/T\,,
\ee
 and the upper (lower) sign applies if $a>0$ ($a<0$). In this form, we can immediately see that the asymptotic solution $y=1$ for $x\ll1$ (which corresponds to the expected $T_\chi=T$ for large $T$) is stable against small perturbations if and only if $a>0$, i.e.~$c>0$. In fact, from the phaseplot shown in Fig.~\ref{fig_phaseplot} it becomes apparent that in this case $\chi$ is kept in thermal equilibrium extremely efficiently for $x\gtrsim10$ and that therefore the evolution of $T_\chi$ actually follows an attractor solution to Eq.~\kref{process}. Once the temperature falls below $x\sim10$, decoupling occurs on a rather short timescale and we observe the expected $T_\chi\propto T^2$ due to the Hubble expansion.

To summarize the results of this Appendix, we have developed a physical understanding for the functional form of the collision term \kref{cgeneral} and found that $T_\chi$ evolves as a function of $T$ according to
\be
  \label{TCDM}
    T_\chi=T\left\{1-\frac{z^{1/(n+2)}}{n+2} \exp[z]\, \Gamma\left[-(n+2)^{-1},z\right]\right\}_{z=\frac{a}{n+2}\left(\frac{T}{M_\chi}\right)^{n+2}} \,,
\ee
essentially independent of the initial conditions. For convenience, we state here again the full expression for the constant $a$ that appears above:
\be
\label{adef}
  a\equiv\sum_{f}\left(\frac{5}{2(2\pi)^9g_\mathrm{eff}}\right)^{1/2}g_\mathrm{SM}c_nN^\pm_{n+3}\,\frac{M_\mathrm{Pl}}{M_\chi}\,,
\ee
where the sum runs over all relativistic scattering partners $f$ (with $g_\mathrm{SM}$ internal degrees of freedom) and $c_n$ is the corresponding scattering amplitude at zero momentum transfer; $g_\mathrm{eff}$ is the effective number of  heat-bath degrees of freedom close to decoupling and $N^\pm_i$ is defined in Eqs.~(\ref{Np}, \ref{Nm}).
Finally, we may now make use of the asymptotic behaviour, Eqs.~(\ref{asymptotics1}, \ref{asymptotics2}), of our general solution to \emph{define the kinetic decoupling temperature} by equating these two limits:
\be
  \label{tdec}
  \frac{T_\mathrm{kd}}{M_\chi}\equiv \left(\left(\frac{a}{n+2}\right)^{1/(n+2)}\Gamma\left[\frac{n+1}{n+2}\right]\right)^{-1}\,.
\ee

\section{The collision term for two-body scattering processes}
\label{app_collint}

In this Appendix, we provide the details for the calculation of the collision term for scattering processes between a non-relativistic particle $\chi$ and relativistic (usually SM) heat-bath particles. For the former, we use $p^\mu=(E,\mathbf{p})$ to denote ingoing momenta, while for the latter we use $k^\mu=(\omega,\mathbf{k})$; the corresponding outgoing quantities are marked with a tilde. With these notations, the collision term takes the form
\be
  \label{Cfull}
  C=\frac12\int\frac{d^3k}{(2\pi)^32\omega}\int\frac{d^3\tilde k}{(2\pi)^32\tilde \omega}\int\frac{d^3\tilde p}{(2\pi)^32\tilde E}(2\pi)^4\delta^{(4)}(\tilde p+\tilde k-p-k)\left|\mathcal{M}\right|^2J\,,
\ee
where $\mathcal{M}$ is the scattering amplitude, summed over final and averaged over initial spin states, and
\be
 \label{Jdef}
 J\equiv g_\mathrm{SM}\left[(1\mp g^\pm)(\omega)\, g^\pm(\tilde\omega)f(\mathbf{\tilde p})-(1\mp g^\pm)(\tilde\omega)\, g^\pm(\omega)f(\mathbf{p})\right]\,.
\ee
We will assume that the relativistic SM particles are thermally distributed, with $g_\mathrm{SM}$ being the number of spin states,
so their phase-space number density is given by 
\be
  g_\mathrm{SM}g^\pm(\omega) = g_\mathrm{SM}\left(e^{\omega/T}\pm1\right)^{-1}
\ee
for fermions (upper signs) and bosons (lower signs), respectively. The only assumption that we will make about the $\chi$ distribution function $f(\mathbf{p})$ is that we can neglect Pauli suppression  factors -- as we have done in \kref{Jdef}. In the context that we are considering here, this is satisfied to a very good approximation.

Since, for kinematical reasons, the average momentum transferred during the scattering events is small, we can expand \kref{Cfull} as 
\bea
   \label{Cexpansion}
   C(E)&=& \sum_{j=0}^{\infty}C^j\,,\\
   C^j&\equiv&\frac12\int\frac{d^3k}{(2\pi)^32\omega}\int\frac{d^3\tilde k}{(2\pi)^32\tilde \omega}\int\frac{d^3\tilde p}{(2\pi)^32\tilde E}\nonumber\\
     &&\times(2\pi)^4\delta(\tilde E+\tilde\omega-E-\omega)\left|\mathcal{M}\right|^2J\left[\frac{1}{j!}D_\mathbf{q}^j(\mathbf{\tilde p})\delta^{(3)}(\mathbf{\tilde p}-\mathbf{p})\right]\,,
\eea
where we have introduced 
\be
  D_\mathbf{q}(\mathbf{\tilde p}) \equiv \mathbf{q}\cdot\nabla_\mathbf{\tilde p}\equiv(\mathbf{\tilde k}-\mathbf{k})\cdot\nabla_\mathbf{\tilde p}\,,
\ee
and the derivatives of the delta function that appear above are as usual defined in terms of integration by parts.

After these preliminaries, let us now start to calculate the expansion coefficients $C^j$. First, we note that $\left.J\right|_{{\mathbf{\tilde p}=\mathbf{p},\,\tilde\omega=\omega}}=0$. From this, it immediately follows that
\be
 C^0=0\,.
\ee
For the next coefficient, we find
\bea
    C^1 &=& -\pi\int\frac{d^3k}{(2\pi)^32\omega}\int\frac{d^3\tilde k}{(2\pi)^32\tilde \omega}\mathbf{q}\cdot\nabla_\mathbf{\tilde p}\left[\frac{\left|\mathcal{M}\right|^2J}{2\tilde E}\delta(\tilde E+\tilde\omega-E-\omega)\right]_{\mathbf{\tilde p}=\mathbf{p}}\\
     &=& -\frac\pi2\int\frac{d^3k}{(2\pi)^32\omega}\int\frac{d^3\tilde k}{(2\pi)^32\tilde \omega}\nonumber\\
     &&{} \times\delta(\tilde\omega-\omega)\frac{\left|\mathcal{M}\right|^2_{t=0}}{E}\left[\mathbf{q}\cdot\nabla_\mathbf{\tilde p}J -\frac{\mathbf{q}\cdot\mathbf{p}}{E}\partial_{\tilde \omega}J\right]_{\mathbf{\tilde p}=\mathbf{p}}\,,
\eea
where we have used
\be
\mathbf{q}\cdot\nabla_\mathbf{\tilde p}\,\delta(\tilde E+\tilde\omega-E-\omega)  = \frac{\mathbf{q}\cdot\mathbf{\tilde p}}{\tilde E}\partial_{\tilde \omega}\delta(\tilde E+\tilde\omega-E-\omega)
\ee
to perform the necessary integration by parts in $\tilde \omega$. Let us now write the scattering amplitude as a function of the Mandelstam variables $s$ and $t$, where $s=M^2_\chi+2\omega(E-\left|\mathbf{p}\right|\cos\theta)$, and expand it in $\left|\mathbf{p}\right|/E$:
\be
  \left|\mathcal{M}\right|^2_{t=0}(s)=\;\mathop{\hspace{-1.5ex}\left|\mathcal{M}\right|^2_{t=0}}_{\hspace{4ex}\theta=\pi/2}-\,\omega\frac{\left|\mathbf{p}\right|}{E}\cos\theta\,\left(\partial_\omega\mathop{\hspace{-1.5ex}\left|\mathcal{M}\right|^2_{t=0}}_{\hspace{4ex}\theta=\pi/2}\right)+\mathcal{O}\left(\frac{\mathbf{p}^2}{E^2}\right)\,.
\ee
With $\mathbf{l}\parallel\mathbf{p}\parallel\nabla_\mathbf{\tilde p}\left.J\right|_{\mathbf{\tilde p}=\mathbf{p}}$, the angular integrals that appear in the above expression for $C^1$ are then of the form
\be
  \int d\Omega\int d\tilde\Omega\, \left|\mathcal{M}\right|^2_{t=0} (\mathbf{q}\cdot\mathbf{l})=
    \frac{16}{3}\pi^2\omega^2E^{-1}(\mathbf{p\cdot l})\left\{\left(\partial_\omega\mathop{\hspace{-1.5ex}\left|\mathcal{M}\right|^2_{t=0}}_{\hspace{4ex}\theta=\pi/2}\right)+\mathcal{O}\left(\frac{\mathbf{p}^2}{E^2}\right)\right\}\,.
\ee
To leading order in $\mathbf{p}^2/E^2$, and with the help  of the useful relation
\be
  g^\pm(1\mp g^\pm)(\omega) = \frac{e^{\omega/T}}{\left(e^{\omega/T}\pm1\right)^2}=-T\partial_\omega g^\pm(\omega)\,,
\ee
we thus arrive at the following result for $C^1$:
\be
  C^1=\frac{g_\mathrm{SM}}{12(2\pi)^3}M_\chi^{-3}\int d\omega\,\omega^4\left(\partial_\omega g^\pm\right)\left(\partial_\omega\mathop{\hspace{-1.5ex}\left|\mathcal{M}\right|^2_{t=0}}_{\hspace{4ex}\theta=\pi/2}\right)\left[TM_\chi\mathbf{p\cdot}\nabla_\mathbf{p} +\mathbf{p}^2\right]\,f(\mathbf{p})\,.
\ee

Let us now compute the next coefficient of the expansion \kref{Cexpansion}:
\bea
  C^2 &=& \frac\pi2\int\frac{d^3k}{(2\pi)^32\omega}\int\frac{d^3\tilde k}{(2\pi)^32\tilde \omega}\left(\mathbf{q}\cdot\nabla_\mathbf{\tilde p}\right)^2\left[\frac{\left|\mathcal{M}\right|^2J}{2\tilde E}\delta(\tilde E+\tilde\omega-E-\omega)\right]_{\mathbf{\tilde p}=\mathbf{p}}\\
 &=& \frac{\pi}{4}\int\frac{d^3k}{(2\pi)^32\omega}\int\frac{d^3\tilde k}{(2\pi)^32\tilde \omega}\left|\mathcal{M}\right|^2_{t=0}\nonumber\\
  &&{ } \times\left[\left(\frac{(\mathbf{q}\cdot\nabla_\mathbf{\tilde p})^2J}{E}-2\frac{(\mathbf{\mathbf{q}\cdot p})(\mathbf{q}\cdot\nabla_\mathbf{\tilde p})J}{E^3}\right)\delta(\tilde\omega-\omega)+\frac{(\mathbf{\mathbf{q}\cdot p})^2}{E^3}J\partial^2_{\tilde \omega}\delta(\tilde\omega-\omega)\right.\nonumber\\
   &&{ } \quad+\left.\left(2\frac{(\mathbf{\mathbf{q}\cdot p})(\mathbf{q}\cdot\nabla_\mathbf{\tilde p})J}{E^2}+\frac{\mathbf{q}^2}{E^2}J-3\frac{(\mathbf{\mathbf{q}\cdot p})^2}{E^4}J\right)\partial_{\tilde \omega}\delta(\tilde\omega-\omega)\right]_{\mathbf{\tilde p}=\mathbf{p}}
\eea
This time, the angular integrals are  of the form
\be
  \int d\Omega\int d\tilde\Omega\, \left|\mathcal{M}\right|^2_{t=0}\,q^iq^j=\frac{16}{3}\pi^2\left(\omega^2+\tilde\omega^2\right)\mathop{\hspace{-1.5ex}\left|\mathcal{M}\right|^2_{t=0}}_{\hspace{4ex}\theta=\pi/2} g^{ij}+\mathcal{O}\left(\frac{\mathbf{p}^2}{E^2}\right)\,.
\ee
Again, we will only keep the leading order terms and, after similar steps as for the computation of $C^1$, we arrive at:
\bea
   C^2 &=& \frac{g_\mathrm{SM}}{12(2\pi)^3} M_\chi^{-3}\int d\omega\,\omega^4\mathop{\hspace{-1.5ex}\left|\mathcal{M}\right|^2_{t=0}}_{\hspace{4ex}\theta=\pi/2}\nonumber\\
   &&\times\left[-TM_\chi^2\partial_\omega g^\pm\Delta_\mathbf{p} +\left(4T\omega^{-1}\partial_\omega g^\pm-\partial_\omega g^\pm+T\partial^2_\omega g^\pm\right)\mathbf{p}\cdot\nabla_\mathbf{p}\right.\nonumber\\
   &&\qquad\left.-3\partial_\omega g^\pm+M_\chi^{-1}\left(4\omega^{-1}\partial_\omega g^\pm+\partial^2_\omega g^\pm\right)\right]\,f(\mathbf{p})\,.
\eea

The terms $C^j$ for $j>2$ only contribute higher-order corrections in $\mathbf{p}^2/E^2$ and $\omega/M_\chi$. To lowest non-vanishing order, and after the necessary partial integrations,  we thus find the following expression for the collision integral:
\bea
  C&=& C^1+C^2\\
    &=& \frac{g_\mathrm{SM}}{12(2\pi)^3} M_\chi^{-2}\int d\omega\,g^\pm(\omega)\partial_\omega\left(\omega^4\mathop{\hspace{-7ex}\left|\mathcal{M}\right|^2_{t=0}}_{\hspace{4.5ex}s=M_\chi^2+2M_\chi\omega}\right)\nonumber\\
    &&\times \Big[M_\chi T\Delta_\mathbf{p} +\mathbf{p}\cdot\nabla_\mathbf{p}+3\Big]\,f(\mathbf{p})\label{Cfullfinal}\eea
Unless one encounters $s$-channel resonances (see \ref{app_LKPscatter} for a treatment of this case), one may now expand the amplitude at zero momentum transfer in $\omega/M_\chi$:
\be
  \label{ndef}
  \mathop{\hspace{-7ex}\left|\mathcal{M}\right|^2_{t=0}}_{\hspace{4.5ex}s=M_\chi^2+2M_\chi\omega}\equiv c_n \left(\frac{\omega}{M_\chi}\right)^n+\mathcal{O}\left(\left(\frac{\omega}{M_\chi}\right)^{n+1}\right)\,.
\ee
Keeping only the lowest order term of this expansion, even the integral over the photon energies can be performed analytically and we arrive, finally, at the following expression:
\be
  \label{Cfinal}
   C=\frac{g_\mathrm{SM}}{12(2\pi)^3} M_\chi^2c_nN^\pm_{n+3}\left(\frac{T}{M_\chi}\right)^{n+4}\Big[M_\chi T\Delta_\mathbf{p} +\mathbf{p}\cdot\nabla_\mathbf{p}+3\Big]\,f(\mathbf{p})\,,
\ee
where (for $j>0$)
\bea
  N_j^+&\equiv& \frac{j+1}{T^{j+1}}\int d\omega\, \omega^jg^+(\omega)=\left(1-2^{-j}\right)\,(j+1)!\,\zeta(j+1)\,,\label{Np}\\
  N_j^-&\equiv& \frac{j+1}{T^{j+1}}\int d\omega\, \omega^jg^-(\omega)=(j+1)!\,\zeta(j+1)\,.\label{Nm}
\eea
Note that in \kref{Cfinal} -- as in all corresponding formulas before -- a summation over all SM scattering partners is understood.

The expression \kref{Cfinal} is the central result of this Appendix;
let us conclude by remarking that the formalism which we have presented here in some detail can straight-forwardly be used to calculate even higher-order corrections to $C$. However, as we have learned in \ref{app_kd}, $\chi$ will be kept in thermal equilibrium until $\mathbf{p}^2/E^2,\langle\omega\rangle/M_\chi\sim10^{-6}-10^{-5}$, so the approximation to keep only leading order expressions in these quantities guarantees a result that is correct to a very high accuracy.

\section{The LKP scattering amplitude}
\label{app_LKPscatter}

\begin{figure}
    \begin{center}
      \includegraphics[width=0.7\textwidth]{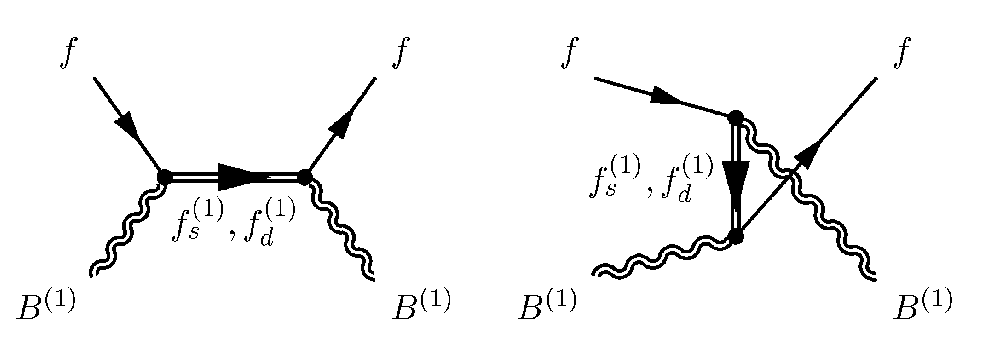}
    \end{center}
     \caption{Feynman diagrams for the elastic scattering of the LKP with SM fermions. In principle, there is also a contribution from a $t$-channel Higgs-exchange, but this is heavily suppressed due to the small Yukawa couplings involved.}
   \label{fig_feynman}
\end{figure}

In this Appendix, we present the LKP scattering amplitude close to kinetic decoupling, which is dominated by the scattering with light SM fermions. The relevant Feynman diagrams are shown in Fig.~\ref{fig_feynman}. Note that in the UED model, at each given KK level, the number of fermions is doubled as compared to the SM, i.e.~singlet and doublet states come each equipped with a whole KK tower of their own. The corresponding couplings of the LKP, well approximated by the $B^{(1)}$, are given by (see, e.g., \cite{Bergstrom:2004nr}):
\be
  \label{scoupling}
  -g_Y\frac{Y_s}{2}\B_\mu\bar f_s^{(1)}\gamma^\mu(1+\gamma^5)f^{(0)}\,+c.c.
\ee
for KK fermion singlet states, $f_s^{(1)}$, and
\be
  \label{dcoupling}
  g_Y\frac{Y_d}{2}\B_\mu\bar f_d^{(1)}\gamma^\mu(1-\gamma^5)f^{(0)}\,+c.c
\ee
for  KK fermion doublet states, $f_d^{(1)}$. Given these couplings, and after summing over all initial states, it is straight-forward to calculate the amplitude at zero momentum transfer (in the limit of $m_f$=0):
\bea
  \label{fullLKPamp}
  \mathop{\hspace{-7ex}\left|\mathcal{M}\right|^2_{t=0}}_{\hspace{4.5ex}s=M_\chi^2+2M_\chi\omega}= \frac{16}{3}g_Y^4Y_s^2\omega^2\frac{M_\B^4\left(M_{f^{(1)}_s}^2-M_\B^2\right)^2+2\omega^2M_\B^2\left(M_{f^{(1)}_s}^4+2M_\B^4\right)}{\left(M_{f^{(1)}_s}^2-M_\B^2+2\omega\right)^2\left(M_\B^2\Gamma^2+\left(M_{f^{(1)}_s}^2-M_\B^2-2\omega\right)^2\right)}\nonumber\\
  \qquad+\left(s\leftrightarrow d\right)\,,
\eea
where $\Gamma$ is the decay width of the corresponding $s$-channel state.
For fermion energies much smaller than the mass difference between the LKP and the KK leptons, $\omega/M_\B\ll\delta\equiv\left(M_{f^{(1)}}-M_\B\right)/M_\B$, we may expand the scattering amplitude as we did in \kref{ndef}:
\bea
  \mathop{\hspace{-7ex}\left|\mathcal{M}\right|^2_{t=0}}_{\hspace{4.5ex}s=M_\chi^2+2M_\chi\omega}&\approx& c_2\left(\frac{\omega}{M_\B}\right)^2\,,\\
  c_2 &\equiv& \frac{16}{3}g_Y^4\left\{Y_d^4\left(\frac{M^2_{B^{(1)}}}{M^2_{f_d^{(1)}}-M^2_{B^{(1)}}}\right)^2 + Y_s^4\left(\frac{M^2_{B^{(1)}}}{M^2_{f_s^{(1)}}-M^2_{B^{(1)}}}\right)^2\right\}\,.\label{approxLKPamp}
\eea

\begin{figure}
    \begin{center}
       \psfrag{x}[t][][1]{$\omega/M_\B$}
       \psfrag{y}[b][][1]{$\left|\mathcal{M}\right|^2_{t=0,s=M_\chi^2+2M_\chi\omega}$}
     \includegraphics[width=0.7\textwidth]{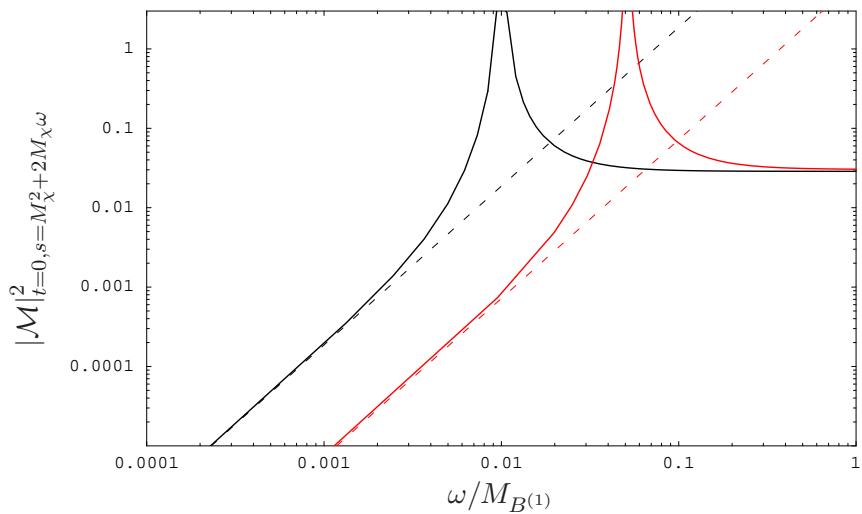}
    \end{center}
     \caption{The full LKP scattering amplitude (solid lines) vs. the approximative expression that scales with the photon energy as $\omega^2$ (dashed lines). For simplicity, we have shown the case of electron scattering, and only taken into account the contribution from left-handed states. The two situations correspond to a mass splitting $\delta\equiv\left(M_{f_s^{(1)}}-M_\B\right)/M_\B$ of 1\% (left) and 5\% (right), respectively.}
   \label{fig_LKP_resonance1}
\end{figure}

In Fig.~\ref{fig_LKP_resonance1}, we compare the full expression for the scattering amplitude with the approximation of keeping only the lowest order term in $\omega/M_\B$; we can see that the low-energy approximation is valid as long as $\omega\lesssim0.1\delta M_\B$, and then starts to deviate considerably. Considering that we expect decoupling temperatures of $T_\mathrm{kd}\sim10\,$MeV (see Section \ref{subsec_LKP}), this condition is easily satisfied for the \emph{average} photon energy, $\langle\omega\rangle=2.70 T$, during the decoupling process. Still, one might be worried about photons in the \emph{tail} of the thermal distribution, which may very well hit the resonance and thereby enhance the effective scattering amplitude considerably.

To address this question, we have to go back to the full expression \kref{Cfullfinal} for the collision integral that we derived in \ref{app_collint}. Going through similar steps as those leading to \kref{process}, we then find that the  process equation for $T_\chi$ to be solved is
\be
   \label{processfull}
   T_{\chi}'-\left[2+a(T)\right]\,T_\chi/T=-a(T),
\ee
where 
\be
 a(T)=\sum_f\left(\frac{5}{2(2\pi)^9g_\mathrm{eff}}\right)^{1/2}g_\mathrm{SM}\frac{M_\mathrm{Pl}}{M_\B^3}T^{-2}\int d\omega\,g^+(\omega)\partial_\omega\left(\omega^4\mathop{\hspace{-7ex}\left|\mathcal{M}\right|^2_{t=0}}_{\hspace{4.5ex}s=M_\chi^2+2M_\chi\omega}\right)
\ee
and the sum runs over all relativistic scattering partners. We may now compare $a(T)$, for $T\sim T_\mathrm{kd}$, between the two situations where we either take the full expression \kref{fullLKPamp} or the approximate expression \kref{approxLKPamp} for the scattering amplitude. Using
\be
  \Gamma=\frac{g_Y^2}{8\pi}\delta^2M_\B,
\ee
which is valid in the limit of small mass splittings $\delta$, we then find that for
\be
  \delta\gtrsim 2.6 \times10^{-3}
\ee
the mistake in $a(T)$ is less than 5\% -- which, from \kref{tdec}, roughly, corresponds to a 1\% error in the estimate for $T_\mathrm{kd}$. For smaller $\delta$, on the other hand, the two ways to calculate $a(t)$ soon start to give grossly different results, rendering this type of error estimation invalid; in such a situation, one has to (numerically) study the asymptotic solution to the full process equation \kref{processfull} in order to determine the decoupling scale. 
This shows that, indead, the high-energy tail of the heat-bath fermions plays an important role in such a situation.

Recalling that in the mUED model we have $\delta\sim10^{-2}$, we will here not further analyze the case of a very pronounced $s$-channel resonance. However, we note that mass splittings of the order of $\delta\sim10^{-3}$ (and smaller) are actually feasible in, e.g.,  non-minimal UED versions. Here, we have provided all the necessary tools to correctly treat such situations.


\end{document}